\documentclass{emulateapj}
\usepackage{apjfonts}
\journalinfo{Accepted for publication in the Astrophysical Journal}
\slugcomment{Submitted to ApJ December 15, 2009; accepted March 4, 2010}
\newcommand{\Msol}{\ensuremath{M_\odot}}

\newcommand{\etal}{\emph{et al.}}
\newcommand{\aanda}{A\&A}
\newcommand{\astroph}[1]{\texttt{astro-ph/{#1}}}
\newcommand{\arxiv}[1]{\texttt{arXiv:{#1}}}
\newcommand{\champagne}{SNLS-03D3bb}

\newcommand{\nickel}{{\ensuremath{^{56}\mathrm{Ni}}}}
\newcommand{\cobalt}{{\ensuremath{^{56}\mathrm{Co}}}}

\newcommand{\mch}{\ensuremath{M_\mathrm{Ch}}}

\begin{document}

% \begin{titlepage}
% \bigskip
% \begin{center}
% Ignore this page
% \end{center}
% \setcounter{page}{0}
% \end{titlepage}

% ============================================================================

% \begin{center}

% {Note:  09-01, written 2009/01/17}

% \bigskip

\title{Nearby Supernova Factory Observations of SN~2007if: \\
       First Total Mass Measurement of a Super-Chandrasekhar-Mass Progenitor}

\author
{
   \mbox{R.~A.~Scalzo},\altaffilmark{1}
   \mbox{G.~Aldering},\altaffilmark{2}
   \mbox{P.~Antilogus},\altaffilmark{3}
   \mbox{C.~Aragon},\altaffilmark{2} 
   \mbox{S.~Bailey},\altaffilmark{3} 
   \mbox{C.~Baltay},\altaffilmark{1} 
   \mbox{S.~Bongard},\altaffilmark{3} 
   \mbox{C.~Buton},\altaffilmark{4} 
   \mbox{M.~Childress},\altaffilmark{2} 
   \mbox{N.~Chotard},\altaffilmark{4} 
   \mbox{Y.~Copin},\altaffilmark{4} 
   \mbox{H.~K.~Fakhouri},\altaffilmark{2} 
   \mbox{A.~Gal-Yam},\altaffilmark{5} 
   \mbox{E.~Gangler},\altaffilmark{4} 
   \mbox{S.~Hoyer},\altaffilmark{1,6} 
   \mbox{M.~Kasliwal},\altaffilmark{7} 
   \mbox{S.~Loken},\altaffilmark{2} 
   \mbox{P.~Nugent},\altaffilmark{8} 
   \mbox{R.~Pain},\altaffilmark{3} 
   \mbox{E.~P\'econtal},\altaffilmark{9} 
   \mbox{R.~Pereira},\altaffilmark{4} 
   \mbox{S.~Perlmutter},\altaffilmark{2} 
   \mbox{D.~Rabinowitz},\altaffilmark{1} 
   \mbox{A.~Rau},\altaffilmark{7} 
   \mbox{G.~Rigaudier},\altaffilmark{9} 
   \mbox{K.~Runge},\altaffilmark{2} 
   \mbox{G.~Smadja},\altaffilmark{4} 
   \mbox{C.~Tao},\altaffilmark{10} 
   \mbox{R.~C.~Thomas},\altaffilmark{8} 
   \mbox{B.~Weaver},\altaffilmark{11,12}
   and \mbox{C.~Wu}\altaffilmark{3} 
}
\affil{}
\email{richard.scalzo@yale.edu}

% Scalzo, Baltay, Bauer, Hoyer, Rabinowitz
\altaffiltext{ 1}{Department of Physics, Yale University,
                  New Haven, CT 06520-8121}
% Aldering, Aragon, Childress, Loken, Perlmutter, Runge
\altaffiltext{ 2}{Physics Division, Lawrence Berkeley National Laboratory,
                  1 Cyclotron Road, Berkeley, CA 94720, USA}
% Antilogus, Bailey, Bongard, Pain, Pereira, Wu
\altaffiltext{ 3}{Laboratoire de Physique Nucl\'eaire et des Hautes Energies,
                  Universit\'e Pierre et Marie Curie,
                  Universit\'e Paris Diderot,
                  CNRS-IN2P3,
                  4 Place Jussieu Tour 43, Rez de Chauss\'ee,
                  75252 Paris Cedex 05, France}
% Birchall
% Present address:  Subaru Telescope, National Astronomical Observatory
%    of Japan, 650 N. Aohoku Pl., Hilo HI 96720
% \altaffiltext{ 4}{Institute for Astronomy, University of Hawaii,
%                   640 N. Aohoku Pl., Hilo, HI 96720}
% Buton, Chotard, Copin, Gangler, Smadja
\altaffiltext{ 4}{Universit\'e de Lyon, 69622, France; 
                  Universit\'e de Lyon 1, France;
                  CNRS/IN2P3, 
                  Institut de Physique Nucl\'eaire de Lyon, France}
% Gal-Yam
\altaffiltext{ 5}{Benoziyo Center for Astrophysics,
                  Weizmann Institute of Science, 76100 Rehovot, Israel}
% Hoyer
\altaffiltext{ 6}{Department of Astronomy, Universidad de Chile,
                  Casilla 36-D, Santiago, Chile}
% Kasliwal, Rau
\altaffiltext{ 7}{Department of Astronomy, California Institute of Technology,
                  1200 E. California Boulevard, MS 105-24,
                  Pasadena, CA 91125, USA}
% Nugent, Thomas
\altaffiltext{ 8}{Computational Cosmology Center,
                  Lawrence Berkeley National Laboratory,
                  1 Cyclotron Road, Berkeley, CA, 94720, USA}
% Pecontal, Rigaudier
\altaffiltext{ 9}{Centre de Recherche Astronomique de Lyon,
                  9 Avenue Charles Andr\'e, 69561 Saint Genis Laval Cedex,
                  France}
% Tao
\altaffiltext{10}{Centre de Physique des Particules de Marseille,
                  163 Av. Luminy, 13288 Marseille Cedex 09, France}
% Weaver
\altaffiltext{11}{University of California, Space Sciences Laboratory,
                  Berkeley, CA 94720}
\altaffiltext{12}{Present address:  New York University,
                  Center for Cosmology and Particle Physics,
                  4 Washington Place, New York, NY 10003, USA}

\shorttitle{SNfactory Observations of SN~2007if}
\shortauthors{Scalzo et al.}
\keywords{supernovae: general;
   supernovae: individual(\objectname{SNe~2003fg, 2006gz, 2007if, 2009dc});
   white dwarfs}

\begin{abstract}
We present photometric and spectroscopic observations of SN~2007if,
an overluminous ($M_V = -20.4$), red ($B-V = 0.16$ at $B$-band maximum),
slow-rising ($t_\mathrm{rise} = 24$~days)
type Ia supernova (SN Ia) in a very faint ($M_g = -14.10$) host galaxy.
A spectrum at 5~days past $B$-band maximum light is a direct match to the
super-Chandrasekhar-mass candidate SN~Ia 2003fg, showing \ion{Si}{2}
and \ion{C}{2} at $\sim 9000$~km~s$^{-1}$.
A high signal-to-noise co-addition of the SN spectral time series reveals
no \ion{Na}{1}~D absorption, suggesting negligible reddening in the host
galaxy, and the late-time color evolution has the same slope as the
Lira relation for normal SNe~Ia.
The ejecta appear to be well mixed, with no strong maximum in $I$-band and
a diversity of iron-peak lines appearing in near-maximum-light spectra.
SN~2007if also displays a plateau in the \ion{Si}{2} velocity extending as
late as $+10$~days, which we interpret as evidence for an overdense shell in
the SN~ejecta.  We calculate the bolometric light curve of the SN and use it
and the \ion{Si}{2} velocity evolution to constrain the mass of the shell and
the underlying SN ejecta, and demonstrate that SN~2007if is strongly
inconsistent with a Chandrasekhar-mass scenario.  Within the context of a
``tamped detonation'' model appropriate for double-degenerate mergers,
and assuming no host extinction, we estimate the total mass of the system
to be $2.4 \pm 0.2$ \Msol, with $1.6 \pm 0.1$ \Msol\ of \nickel\ and with
0.3--0.5 \Msol\ in the form of an envelope of unburned carbon/oxygen.
Our modeling demonstrates that the kinematics of shell entrainment provide
a more efficient mechanism than incomplete nuclear burning for producing
the low velocities typical of super-Chandrasekhar-mass SNe~Ia.
\end{abstract}

% ############################################################################

\section{Introduction}

% \include{background.tex}

% TODO RS:  Rethink punch of paper.  What's the new angle?
% -- Shell subclass nominally staked out by Quimby et al.; see who's cited.
%    Need to mention this in re: impact of super-C findings on cosmology.
% -- Maeda et al.:  Adequately responded to below, not mentioned above
% -- Relative rate stuff needs to have additional SNe to go for it
% Maybe add a short section on SNF 20080723-012?  or a separate letter?

Type Ia supernovae
(SNe~Ia) are of vital importance as luminosity distance indicators for
measuring the expansion history of the Universe
\citep{riess98,scp99}.  % scp03,riess04,mwv07,scpunion,hicken09,kessler09
They have a small dispersion ($\sim 0.35$ mag) in intrinsic peak luminosity,
which can be further reduced to 0.16--0.18 mag by applying a well-known
correction dependent on the width, or decay rate, of the light curve
\citep{phillips99,salt2,mlcs2k2}.
Searches for other luminosity correlates with which
to derive even more accurate luminosity distances from SNe~Ia are underway,
with some methods delivering core Hubble residual dispersions as low as
0.12 mag \citep{sjb09,wang09,csp09}.  SNe~Ia are
generally understood to result from the thermonuclear explosion of at least
one carbon/oxygen white dwarf.  However, the underlying distribution of
SN~Ia progenitor systems and explosions mechanisms, and the relative rates
of possible different physical subclasses of SN~Ia events, remain poorly
constrained, with potential consequences for the average luminosity of
events.  Any systematic effect which may influence the luminosity of 
different SN~Ia subpopulations
at the level of a few percent has become a cause for concern for
next-generation experiments, particularly redshift-dependent effects
\citep{kim04,linder04}.  A better understanding of the progenitor
systems will place these corrections on a much firmer conceptual footing
and place limits on hitherto uncontrolled astrophysical systematics in
SN~Ia luminosity distance measurements.

Rare SN~Ia events displaying extreme characteristics or evidence of
unusual physics can often point the way toward other, less extreme instances
of similar physics which may be lurking in the otherwise undifferentiated
sample of ``normal'' events.  For example, one commonly-invoked rationale
for the uniformity in pre-correction SN~Ia luminosities is that they start
with the same amount of fuel, and are triggered by the same physics:
the SN~Ia progenitor explodes when its mass nears the Chandrasekhar limit,
$\mch = 1.4\ \Msol$, its mass is completely unbound and converted mostly
to heavier elements, especially \nickel, the decay of which powers the
SN~Ia light curve.  The \emph{single-degenerate} scenario \citep[SD;][]{wi73}
ensures that the white dwarf slowly approaches \mch\ via accretion from
a non-degenerate companion.  In contrast, the \emph{double-degenerate}
scenario \citep[DD;][]{it84}, in which two white dwarfs in a binary system
merge and explode, provides a way for SN~Ia progenitors to exceed \mch\
and to give rise to more luminous events.  There are also some arguments
that single-degenerate, differentially rotating white dwarfs with mass
exceeding \mch\ significantly can exist \citep{yl05}, although the
inclusion of baroclinic and magnetohydrodynamic instabilities appears to
preclude this \cite[e.g.,][]{piro08}.  There may therefore be a
population of SNe~Ia with a distribution of masses greater than \mch,
with different explosion physics that interferes with luminosity
standardization.  The relative rate of such events among SNe~Ia in general
may also depend on redshift, and unless they can be identified or their
luminosities accurately calibrated, they need not be common to produce
significant biases in reconstructions of the dark energy equation of state.

There are at least four documented examples of overluminous
SN~Ia explosions with progenitor mass potentially exceeding \mch.
The first known example of the class was SN~2003fg
\citep[\champagne;][]{howell06}; SN~2006gz \citep{hicken07},
SN~2007if \citep{cbet2007if}, and SN~2009dc
\citep{tanaka09,yamanaka09,silverman10}
were discovered later as events spectroscopically similar to SN 2003fg.
The main evidence cited for a very massive progenitor in each of these cases
was the extremely high luminosity of each of these events, and by
inference unusually large \nickel\ synthesis.
SNe~2003fg, and later SNe 2006gz and 2009dc, were noted for being overluminous
($M_V \sim -20$), with unusually wide light curves (``stretch'' $s > 1.1$)
and \ion{C}{2} lines, evidence for unburned carbon, in early or near-maximum
spectra.  In SN~2003fg, narrow, low-velocity
($8000$--$9000$~km~s$^{-1}$) \ion{Si}{2} lines near maximum light
have been interpreted as evidence for a high gravitational binding
energy, further supporting the hypothesis of a very massive progenitor;
low velocities were also found in SN~2009dc.
Ejecta velocities inferred from spectra of SN~2006gz were closer to
those of normal SNe~Ia.

SN~2007if was discovered by the Texas Supernova Search \citep{cbet2007if}
in unfiltered ROTSE-IIIb images taken on 2007~August~16.3 UT.
It was found independently as SNF20070825-001 by the Nearby Supernova
Factory \citep[SNfactory;][]{snf}, in an image taken in a red (RG610) filter
with the QUEST-II camera \citep{baltay07} on the Palomar Observatory
Oschin 1.2~m Schmidt telescope (``Palomar-QUEST'') on 2007~August~25.4.
No host galaxy was visible in the discovery images; nor in any available
sky survey images, including the Sloan Digital Sky Survey (SDSS),
POSS and USNO, making the redshift
determination and interpretation of early-phase spectra uncertain.
An optical spectrum we obtained with the SuperNova Integral Field Spectrograph
\citep[SNIFS;][]{snifs}
on the University of Hawaii 2.2~m telescope on 2007~August~26.5 UT revealed a
blue continuum not obviously like a type~Ia supernova.  Spectroscopy taken
at the Hobby-Eberly Telescope on 2007~August~29 also failed to identify the
nature of the event \citep{cbet2007if}.
However, a later spectrum we obtained with the Double Spectrograph on the Hale
5-m telescope at Palomar, on 2007~September~6.5 UT, identified SN~2007if
as a SN~Ia, apparently well before peak.
A further SNIFS spectrum taken on September~10.5 UT turned out to be an
unambiguous match to a published spectrum of SN~2003fg \citep{howell06}.
Cross-correlation of the September~10.5 SNIFS spectrum with the SNLS spectrum
of SN~2003fg suggested a redshift of $0.070 \pm 0.005$.

The faintness of the host, coupled with the unusually
large luminosity of the supernova, presented a challenge for the detection
of line emission from the host as late as a full year after explosion.
More recently, however, an optical spectrum of the host galaxy was obtained
on 2009~August~24.5 with the Low Resolution Imaging Spectrograph (LRIS)
at the Keck Telescope on Mauna Kea, showing
H$\alpha$ and \ion{O}{3}~$\lambda 3727$ at a heliocentric redshift of
$0.07416 \pm 0.00082$.  This new redshift measurement allows more accurate
determination of the SN luminosity and the ejecta velocity scale, which in
turn enables a measurement of the total mass in the explosion.

In Sections 2 and 3 we present our detailed photometric and spectroscopic
observations of SN~2007if and its host galaxy.  In Section 4 we present the
bolometric light curve of the SN and an estimate of the mass of \nickel\
synthesized in the explosion, which we find to nominally exceed \mch.
In Section 5 we argue that the red color of the SN, its unusually long
(24-day) rise time, and the existence of a plateau in the inferred
photospheric velocity are best explained by the existence of an overdense
shell in the ejecta, probably caused by the entrainment of an unburned
carbon-oxygen envelope.  We also estimate the total mass ejected in the SN,
and the fraction of that mass residing in the shell and envelope.
Since SN~Ia mass estimates are often sensitive to the assumed kinetic energy
of the explosion, we consider in Section 6 the importance of shell structure
on arguments associated with mass estimates in the literature, and ask whether
shell structure may be more common in super-Chandrasekhar-mass SN~Ia
candidates than previously believed.  We summarize and conclude in Section 7.

% ############################################################################

\section{Observations}

% ============================================================================

\subsection{Optical Spectroscopy}
\label{sec:spectra_obs}

Observations of SN~2007if were obtained with SNIFS \citep{snf,snifs},
operated by the SNfactory;
the observing log is shown in Table~\ref{tbl:speclog}.
SNIFS is a fully integrated instrument optimized
for automated observation of point sources on a diffuse background
over the full optical window at moderate spectral resolution.  It
consists of a high-throughput wide-band pure-lenslet integral field
spectrograph \citep[IFS, ``\`a la
TIGER'';][]{bacon95,bacon00,bacon01}, a multifilter photometric
channel to image the field surrounding the IFS for atmospheric
transmission monitoring simultaneous with spectroscopy, and an
acquisition/guiding channel.  The IFS possesses a fully filled
$6\farcs 4 \times 6\farcs 4$ spectroscopic field of view (FOV) subdivided
into a grid of $15 \times 15$ spatial elements (spaxels), a
dual-channel spectrograph covering 3200--5200~\AA\ and
5100--10000~\AA\ simultaneously, and an internal calibration unit
(continuum and arc lamps).  SNIFS is continuously mounted on the south
bent Cassegrain port of the University of Hawaii 2.2~m telescope
(Mauna Kea) and is operated remotely.  The SNIFS spectra of SN~2007if
were reduced using our dedicated data reduction procedure, similar to
that presented in Section 4 of \cite{bacon01}.  A brief discussion on the
spectrographic pipeline was presented in \cite{snf2005gj}, here we
outline changes to the pipeline since that work, but leave a complete
discussion of the reduction pipeline to subsequent publications
focused on the instrument itself.

After standard CCD preprocessing and subtraction of a low-amplitude
diffuse-light component, the 225 spectra from the individual spaxels
of each SNIFS exposure were extracted from each blue and red
spectrograph exposure, and re-packed into two
$(x,y,\lambda)$-datacubes. This highly specific extraction is based
upon a detailed optical model of the instrument including
interspectrum crosstalk corrections.  The datacubes were then
wavelength-calibrated, using arc lamp exposures acquired immediately
after the science exposures, and spectro-spatially flat-fielded, using
continuum lamp exposures obtained during the same night. Cosmic rays
were detected and corrected using a three-dimensional-filtering scheme
applied to the datacubes.

SN and standard star spectra were extracted from each
$(x,y,\lambda)$-datacube using three-dimensional point-spread function (PSF)
fit photometry over a uniform background \citep{buton09}. The PSF is modeled
semi-analytically as a constrained sum of a Gaussian (describing the
core) and a Moffat function (simulating the wings). The correlations
between the different shape parameters, as well as their wavelength
dependencies, have been trained on a set of 300~standard star 
observations in various conditions of seeing and telescope focus between
2004 and 2007 with SNIFS. This resulted in a chromatic PSF empirical
model depending only on an effective width (mimicking seeing) and a
flattening parameter (for small imaging defocus and guiding
errors). The three-dimensional PSF fit takes the atmospheric differential
refraction into account without resampling.

During photometric nights, the SN spectra were flux calibrated using a
flux solution and a mean atmospheric extinction derived simultaneously
from all spectrophotometric standard stars observed during the same
night \citep{buton09}. In non-photometric conditions, an effective
attenuation measurement for each exposure was made using the stars
observed by the SNIFS multifilter photometric channel. Objects in the
field (spatially subdivided into five regions each monitoring a
different wavelength range and treated separately) were detected using
\texttt{SExtractor} \citep{bertin96}, and the produced catalogs
matched to (manually-inspected) star catalogs created from deep stacks
of the same field. An adapted version of the Supernova Legacy Survey
photometry code, \texttt{poloka}, was then used to determine the
so-called photometric flux ratio between each exposure and a reference
exposure chosen from a reference night considered to be photometric.
A convolution kernel is computed using the matched objects between each
pair of images in order to make both comparable flux-wise, meaning
that when it is applied to the best quality image (the reference one,
with best seeing) it homogenizes it to the same ``photometric frame''
as the worst image. The photometric flux ratio equals the integral of
the convolution kernel, an estimation of the relative flux extinction in
each filter band for the observations on non-photometric nights, and is used
to correct for attenuation by clouds.  Seeing and the stability of the
atmospheric transmission were assessed using quantitative analyses of
SNIFS guider video frames acquired during our exposures, along with
deglitched CFHT Skyprobe data.

\begin{deluxetable}{rrrc}
\tabletypesize{\footnotesize}
\tablecaption{Observing log for SNIFS spectra of SN~2007if\label{tbl:speclog}}
\tablehead{
   \colhead{MJD\tablenotemark{a}} &
   \colhead{Phase\tablenotemark{b}} &
   \colhead{Exp. Time (s)} & \colhead{Airmass}
}
\startdata
54338.5 & -9.0 & 1200 & 1.004 \\
54353.5 &  4.9 & 1900 & 1.003 \\
54355.5 &  6.8 & 1230 & 1.010 \\
54358.5 &  9.6 &  900 & 1.013 \\
54363.5 & 14.2 & 1740 & 1.005 \\
54365.5 & 16.1 & 1970 & 1.003 \\
54373.4 & 23.5 & 1800 & 1.007 \\
54375.5 & 25.4 & 2000 & 1.004 \\
54378.5 & 28.2 & 1800 & 1.005 \\
54378.5 & 28.2 & 1800 & 1.025 \\
54385.4 & 34.7 & 1800 & 1.005 \\
54390.4 & 39.3 & 1800 & 1.003 \\
54393.4 & 42.1 & 2000 & 1.027 \\
54395.4 & 43.9 & 1800 & 1.034 \\
54395.4 & 43.9 & 1800 & 1.009 \\
54400.4 & 48.6 & 1800 & 1.017 \\
54400.4 & 48.6 & 1800 & 1.052 \\
54403.4 & 51.4 & 1800 & 1.017 \\
54403.4 & 51.4 & 1800 & 1.052 \\
54415.3 & 62.5 & 1800 & 1.052 \\
54420.3 & 67.2 & 2200 & 1.003

\enddata
\tablenotetext{a}{Observer frame JD - 2400000.5.}
\tablenotetext{b}{In rest-frame days relative to $B$-band maximum light.}
\end{deluxetable}

To supplement the ANDICAM imaging photometry discussed in the next section,
we synthesized additional
rest-frame Bessell $BVRI$ photometry from the SNIFS flux-calibrated spectra.
The synthetic photometry is shown alongside the imaging photometry in
Table~\ref{tbl:lc-rest} and in Figures~\ref{fig:lc-bvrij} and
\ref{fig:lc-bvlira}.

% ============================================================================

\subsection{ANDICAM $BVRI$ photometry}
\label{subsec:bvri_obs}

Follow-up $BVRI$ photometry of SN~2007if using ANDICAM on the CTIO 1.3-m
was obtained through the Small and Moderate Aperture Research Telescope
System (SMARTS) Consortium, from 2007~September~10 through 2007~December~20.
Each band was exposed for approximately 240~s.
Further spectroscopic follow-up was also obtained with SNIFS extending
through 2007~November~26.  Since the nature and type of SN~2007if were not
immediately apparent, SMARTS observations began only on 2007~September~10.5,
when the light curve was already in decline.  Final reference images
for subtraction of host galaxy light were obtained in the 2008 observing
season; although the host galaxy is very faint, at late times its contribution
could still be significant.

The SMARTS photometry was processed using an automated pipeline based on
IRAF \citep{iraf}.  SMARTS images were bias-subtracted,
overscan-subtracted, and flat-fielded using \texttt{ccdproc}.  Point sources
were detected (\texttt{daofind}) and their instrumental magnitudes measured
(\texttt{phot}), then aperture-corrected to a 6\arcsec\ aperture (\texttt{mkapfile}).
The images were registered to a standard detector coordinate system
(\texttt{xyxymatch}, \texttt{geomap}, \texttt{gregister}).
The resampled final reference images were combined (\texttt{imsum})
for added depth in each band, rejecting detector defects and cosmic rays
by discarding the brightest and faintest value at each pixel location in
the stack.  An absolute calibration (zeropoint, extinction and color terms)
was established on photometric nights from observations of \cite{landolt92}
standards, fitting a zeropoint and extinction coefficient for each night
separately as well as a color term constant across all nights.
The calibration was transferred to the field stars for each photometric
night separately using the zeropoint and extinction but ignoring the color
terms, producing magnitudes on a ``natural'' ANDICAM system which agrees
with the Landolt system for stars with $B-V = V-R = R-I = 0$.
These calibrated magnitudes were then averaged over photometric nights
to produce final calibrated ANDICAM magnitudes for the field stars.
The final reference co-add was normalized to each image in turn,
with the best-seeing image in each pair convolved to match the worst
(\texttt{fitpsf},\texttt{psfmatch}); the (undetected) host galaxy light was
then removed by subtraction.  The positions of all detections of the SN with
signal-to-noise ratio (S/N) greater than 10 were averaged, and this final
position used to measure the flux in each image, in order to achieve more
accurate photometry for noisy measurements at late times.  Systematic errors
were estimated for the resampling, flux normalization and PSF convolution
steps by measuring the dispersion in the residuals
(from a light curve constant in time)
of the field stars.  We estimate an end-to-end systematic floor of 1\%
on our differential photometry, so that our observations of the SN~are
limited only by sky noise.

To characterize the ANDICAM system throughput as a function of wavelength
in each band, we use a product in each band of the KPNO filter
transmission curve, the quantum efficiency of a Fairchild 447 CCD as
measured by the manufacturer (including MgO$_2$ quartz window transmission),
and the reflectivity of two layers of aluminum.  We adjust the central
wavelength of each ANDICAM passband using SMARTS observations of
spectrophotometric standards \citep{stritz05}, such that our calibrated
natural-system photometry matches synthetic photometry in the adjusted
passbands from the standard star spectra.
After these steps, the SN's ANDICAM-system $BVRI$ light curve was measured
by comparison to field stars.  The observer-frame ANDICAM magnitudes were
corrected for Galactic extinction using $E(B-V) = 0.079$ \citep{sfd}
and $R_V = 3.1$ \citep{cardelli}, then $K$-corrected \citep{nkp02}
to rest-frame Bessell $BVRI$ bandpasses \citep{bessell90}, using
the adjusted ANDICAM filter bandpasses and the SNIFS spectral time series.

% ============================================================================

\subsection{ANDICAM $J$ photometry}
\label{subsec:j_obs}

The SMARTS observations were taken with ANDICAM in dual CCD-IR mode, so that
a $J$ light curve was obtained simultaneously with the $BVRI$ light curve.
Four dithered 50~s $J$-band exposures were taken during each 240~s
optical-band exposure, for a total $J$ exposure time of 800~s each night.

The $J$-band images were reduced using IRAF.
Separate bias and overscan levels for each of the four amplifiers of the
detector were subtracted from the ANDICAM images, which were then
flat-fielded using a superflat made from dome flat images taken nightly.
Bad pixels were identified as $3\sigma$-deviant pixels in the superflat.
Sources in the field were generally detected at low S/N,
precluding automatic registration of single dithered exposures.
We therefore performed an initial co-add of the exposures at each dither
position without further registration; we estimate a positional error of
approximately $0 \farcs 5$, resulting in a modest loss in S/N.
Sources were then detected in the co-add at each dither position, and the
single-dither co-adds were registered to each other to produce a final
co-add for the night.  Host galaxy subtraction was neglected for $J$-band,
since registration was difficult and since the high sky noise in these
measurements dominates over any systematic error from unsubtracted light
from such a faint host.
The field star magnitudes were calibrated using standard star observations
on nights reported as photometric by the SMARTS queue observer,
with extinction corrections based on a seasonal average $J$-band extinction
coefficient \citep[0.1~mag~airmass$^{-1}$;][]{frogel98} for CTIO.
The SN magnitudes were then measured via differential
photometry using three suitable non-variable stars in the FOV,
similar to the process used in reducing the optical data.

While we have no near-infrared spectra of SN~2007if with which to perform
$K$-corrections on the $J$-band magnitudes, we use the spectra of
the 1991T-like SN~1999ee \citep{hamuy02} as a reasonable approximation.
The transfer function used was the product of the quantum efficiency of the
Rockwell HgCdTe ``Hawaii'' array and the $J$-band filter transmission
as measured by the manufacturer.

Despite this somewhat approximate treatment, we expect based on residuals
for the field stars that our systematic errors are of order 10\% or less,
in general comparable to or less than the sky noise.  These are included
in our stated error bars.  % The rest-frame $J$ light curve of SN~2007if
% is shown in Table~\ref{tbl:lc-rest} and in Figure~\ref{fig:lc-bvrij}.

\begin{deluxetable*}{lrcccccc}
\tabletypesize{\footnotesize}
\tablecaption{Rest-frame $BVRIJ$ light curve of SN~2007if \label{tbl:lc-rest}}
\tablehead{
   \colhead{MJD\tablenotemark{a}} &
   \colhead{Phase\tablenotemark{b}} &
   \colhead{$B$} &
   \colhead{$V$} &
   \colhead{$R$} &
   \colhead{$I$} &
   \colhead{$J$} &
   \colhead{Instrument}
}
\startdata
% \sidehead{Quality cuts only:}
% 0 & $0 \pm 0$ & 0 & 0 & 0 & 0 & 0 & 0 \\
54337.3 & -10.3 &              \ldots &              \ldots & $17.78 \pm 0.10$ &              \ldots &              \ldots & QUEST  \\
54338.5 &  -9.2 &              \ldots & $17.67 \pm 0.02$ &              \ldots &              \ldots &              \ldots & SNIFS-P \\
54338.5 &  -9.2 & $17.54 \pm 0.04$ & $17.62 \pm 0.06$ & $17.53 \pm 0.04$ & $17.64 \pm 0.06$ &              \ldots & SNIFS-S \\
54346.3 &  -2.0 &              \ldots &              \ldots & $17.15 \pm 0.07$ &              \ldots &              \ldots & QUEST  \\
54353.4 &   4.6 &              \ldots &              \ldots & $17.29 \pm 0.05$ &              \ldots &              \ldots & QUEST  \\
54353.5 &   4.7 & $17.45 \pm 0.04$ & $17.29 \pm 0.05$ & $17.27 \pm 0.04$ & $17.46 \pm 0.05$ &              \ldots & SNIFS-S \\
54354.4 &   5.5 & $17.49 \pm 0.02$ & $17.28 \pm 0.02$ & $17.28 \pm 0.01$ & $17.40 \pm 0.02$ &              \ldots & SMARTS \\
54354.4 &   5.6 &              \ldots &              \ldots & $17.19 \pm 0.11$ &              \ldots &              \ldots & QUEST  \\
54355.5 &   6.6 &              \ldots & $17.29 \pm 0.02$ &              \ldots &              \ldots &              \ldots & SNIFS-P \\
54355.5 &   6.6 & $17.49 \pm 0.04$ & $17.24 \pm 0.04$ & $17.21 \pm 0.03$ & $17.38 \pm 0.05$ &              \ldots & SNIFS-S \\
54356.3 &   7.3 & $17.57 \pm 0.03$ & $17.27 \pm 0.02$ & $17.29 \pm 0.01$ & $17.43 \pm 0.02$ & $18.41 \pm 0.17$ & SMARTS \\
54358.2 &   9.2 & $17.69 \pm 0.04$ & $17.33 \pm 0.03$ & $17.34 \pm 0.01$ & $17.49 \pm 0.02$ & $18.43 \pm 0.17$ & SMARTS \\
54358.5 &   9.4 &              \ldots & $17.36 \pm 0.02$ &              \ldots &              \ldots &              \ldots & SNIFS-P \\
54358.5 &   9.4 & $17.66 \pm 0.09$ & $17.19 \pm 0.09$ & $17.18 \pm 0.05$ & $17.34 \pm 0.10$ &              \ldots & SNIFS-S \\
54361.3 &  12.0 & $17.96 \pm 0.02$ & $17.50 \pm 0.03$ & $17.47 \pm 0.01$ & $17.58 \pm 0.03$ &              \ldots & SMARTS \\
54363.2 &  13.8 & $18.17 \pm 0.03$ & $17.55 \pm 0.03$ & $17.47 \pm 0.04$ & $17.60 \pm 0.03$ & $19.16 \pm 0.14$ & SMARTS \\
54363.5 &  14.0 &              \ldots & $17.59 \pm 0.04$ &              \ldots &              \ldots &              \ldots & SNIFS-P \\
54363.5 &  14.0 & $18.14 \pm 0.07$ & $17.55 \pm 0.06$ & $17.48 \pm 0.04$ & $17.59 \pm 0.07$ &              \ldots & SNIFS-S \\
54365.5 &  15.9 & $18.29 \pm 0.09$ & $17.55 \pm 0.06$ & $17.46 \pm 0.04$ & $17.52 \pm 0.07$ &              \ldots & SNIFS-S \\
54366.3 &  16.6 & $18.52 \pm 0.06$ & $17.71 \pm 0.05$ & $17.62 \pm 0.05$ & $17.68 \pm 0.11$ &              \ldots & SMARTS \\
54372.3 &  22.2 & $19.05 \pm 0.32$ &              \ldots & $17.68 \pm 0.23$ & $17.55 \pm 0.29$ & $19.08 \pm 0.13$ & SMARTS \\
54373.4 &  23.3 & $19.02 \pm 0.12$ & $18.06 \pm 0.08$ & $17.76 \pm 0.05$ & $17.63 \pm 0.06$ &              \ldots & SNIFS-S \\
54375.2 &  24.9 & $19.38 \pm 0.08$ & $18.22 \pm 0.04$ & $17.85 \pm 0.02$ & $17.73 \pm 0.04$ & $19.17 \pm 0.20$ & SMARTS \\
54375.4 &  25.2 & $19.15 \pm 0.11$ & $18.16 \pm 0.06$ & $17.85 \pm 0.04$ & $17.73 \pm 0.05$ &              \ldots & SNIFS-S \\
54378.2 &  27.8 & $19.49 \pm 0.06$ & $18.37 \pm 0.03$ & $17.98 \pm 0.02$ & $17.85 \pm 0.03$ &              \ldots & SMARTS \\
54378.5 &  28.0 & $19.38 \pm 0.17$ & $18.33 \pm 0.09$ & $17.99 \pm 0.04$ & $17.83 \pm 0.07$ &              \ldots & SNIFS-S \\
54382.3 &  31.5 & $19.65 \pm 0.07$ & $18.46 \pm 0.03$ & $18.11 \pm 0.02$ & $17.85 \pm 0.03$ &              \ldots & SMARTS \\
54385.4 &  34.4 & $19.57 \pm 0.14$ & $18.48 \pm 0.07$ & $18.11 \pm 0.04$ & $17.87 \pm 0.06$ &              \ldots & SNIFS-S \\
54388.1 &  37.0 & $19.81 \pm 0.07$ & $18.62 \pm 0.05$ & $18.25 \pm 0.03$ & $17.99 \pm 0.03$ & $19.16 \pm 0.10$ & SMARTS \\
54388.2 &  37.1 &              \ldots & $18.59 \pm 0.07$ &              \ldots &              \ldots &              \ldots & SNIFS-P \\
54388.4 &  37.3 &              \ldots & $18.64 \pm 0.07$ &              \ldots &              \ldots &              \ldots & SNIFS-P \\
54388.4 &  37.3 & $19.84 \pm 0.43$ & $18.69 \pm 0.22$ & $18.26 \pm 0.10$ & $17.98 \pm 0.14$ &              \ldots & SNIFS-S \\
54390.4 &  39.1 &              \ldots & $18.71 \pm 0.05$ &              \ldots &              \ldots &              \ldots & SNIFS-P \\
54390.4 &  39.1 & $19.70 \pm 0.15$ & $18.66 \pm 0.09$ & $18.31 \pm 0.05$ & $18.05 \pm 0.07$ &              \ldots & SNIFS-S \\
54392.2 &  40.8 & $19.88 \pm 0.10$ & $18.81 \pm 0.04$ & $18.38 \pm 0.03$ & $18.14 \pm 0.04$ &              \ldots & SMARTS \\
54393.3 &  41.8 &              \ldots & $18.81 \pm 0.05$ &              \ldots &              \ldots &              \ldots & SNIFS-P \\
54393.4 &  41.9 & $19.79 \pm 0.26$ & $18.77 \pm 0.14$ & $18.39 \pm 0.07$ & $18.13 \pm 0.10$ &              \ldots & SNIFS-S \\
54395.1 &  43.5 & $19.88 \pm 0.23$ &              \ldots &              \ldots & $18.45 \pm 0.13$ &              \ldots & SMARTS \\
54395.3 &  43.7 & $19.83 \pm 0.18$ & $18.82 \pm 0.10$ & $18.48 \pm 0.06$ & $18.22 \pm 0.07$ &              \ldots & SNIFS-S \\
54400.2 &  48.2 & $20.49 \pm 0.49$ &              \ldots & $18.68 \pm 0.25$ & $18.38 \pm 0.25$ & $20.55 \pm 0.27$ & SMARTS \\
54400.4 &  48.4 &              \ldots & $19.06 \pm 0.09$ &              \ldots &              \ldots &              \ldots & SNIFS-P \\
54400.4 &  48.4 & $19.74 \pm 0.51$ & $19.19 \pm 0.21$ & $18.90 \pm 0.11$ & $18.71 \pm 0.13$ &              \ldots & SNIFS-S \\
54403.4 &  51.2 &              \ldots & $19.09 \pm 0.04$ &              \ldots &              \ldots &              \ldots & SNIFS-P \\
54403.4 &  51.2 & $20.07 \pm 0.51$ & $19.14 \pm 0.22$ & $18.83 \pm 0.13$ & $18.60 \pm 0.16$ &              \ldots & SNIFS-S \\
54404.2 &  52.0 & $20.04 \pm 0.10$ & $19.18 \pm 0.06$ & $18.75 \pm 0.04$ & $18.61 \pm 0.06$ & $20.43 \pm 0.26$ & SMARTS \\
54408.2 &  55.6 & $20.20 \pm 0.10$ & $19.26 \pm 0.05$ & $18.94 \pm 0.04$ & $18.61 \pm 0.05$ & $20.77 \pm 0.33$ & SMARTS \\
54412.3 &  59.5 & $20.20 \pm 0.10$ & $19.36 \pm 0.05$ & $19.08 \pm 0.04$ & $18.97 \pm 0.07$ & $20.99 \pm 0.45$ & SMARTS \\
54415.3 &  62.3 &              \ldots & $19.28 \pm 0.05$ &              \ldots &              \ldots &              \ldots & SNIFS-P \\
54415.3 &  62.3 & $20.31 \pm 0.33$ & $19.43 \pm 0.20$ & $19.19 \pm 0.12$ & $18.98 \pm 0.17$ &              \ldots & SNIFS-S \\
54418.1 &  64.9 & $20.28 \pm 0.10$ & $19.50 \pm 0.05$ & $19.23 \pm 0.04$ & $19.22 \pm 0.07$ &              \ldots & SMARTS \\
54420.3 &  66.9 &              \ldots & $19.50 \pm 0.04$ &              \ldots &              \ldots &              \ldots & SNIFS-P \\
54420.3 &  67.0 & $20.25 \pm 0.22$ & $19.42 \pm 0.14$ & $19.25 \pm 0.09$ & $19.09 \pm 0.12$ &              \ldots & SNIFS-S \\
54422.1 &  68.6 & $20.16 \pm 0.14$ & $19.50 \pm 0.08$ & $19.41 \pm 0.06$ & $19.26 \pm 0.09$ &              \ldots & SMARTS \\
54430.4 &  76.4 &              \ldots & $19.49 \pm 0.10$ &              \ldots &              \ldots &              \ldots & SNIFS-P \\
54431.1 &  77.0 & $20.40 \pm 0.27$ & $19.60 \pm 0.14$ & $19.65 \pm 0.14$ & $19.38 \pm 0.13$ &              \ldots & SMARTS \\
54435.1 &  80.7 & $20.67 \pm 0.17$ & $19.76 \pm 0.07$ & $19.69 \pm 0.07$ & $19.63 \pm 0.15$ &              \ldots & SMARTS \\
54438.1 &  83.5 & $20.70 \pm 0.15$ & $19.73 \pm 0.06$ & $19.80 \pm 0.06$ & $19.74 \pm 0.12$ &              \ldots & SMARTS \\
54442.1 &  87.2 & $20.53 \pm 0.14$ & $19.82 \pm 0.07$ & $19.96 \pm 0.08$ & $20.23 \pm 0.23$ &              \ldots & SMARTS \\
54445.1 &  90.0 & $20.84 \pm 0.18$ & $20.00 \pm 0.08$ & $19.81 \pm 0.07$ & $19.81 \pm 0.14$ &              \ldots & SMARTS \\
54448.1 &  92.8 & $20.87 \pm 0.18$ & $19.92 \pm 0.07$ & $19.98 \pm 0.07$ & $19.88 \pm 0.14$ &              \ldots & SMARTS \\
54451.1 &  95.6 & $20.84 \pm 0.33$ & $19.82 \pm 0.12$ & $20.00 \pm 0.14$ & $19.79 \pm 0.18$ &              \ldots & SMARTS

\enddata
% \tablecomments{...}
\tablenotetext{a}{Observer frame $\mathrm{JD} - 2400000.5$.}
\tablenotetext{b}{In rest-frame days relative to $B$-band maximum light.}
\end{deluxetable*}
% ============================================================================

\subsection{Additional optical photometry}
\label{subsec:plus_obs}

To further constrain the SN's multi-band behavior near maximum light,
the ANDICAM data were supplemented with observations near maximum from the
broad-band RG610 data collected by the Palomar-QUEST supernova search,
and with additional $V$-band photometric observations taken with the SNIFS
imaging channel when adjusting the telescope's pointing to place the SN
into the FOV of the SNIFS integral field unit.
Stars in the RG610 images of the SN were matched to
corresponding $R$-band and $I$-band images taken with ANDICAM,
and their $R$ magnitudes were corrected to a natural Palomar-QUEST
RG610 system using a linear color correction,
$RG610 = R + c(R-I)$ (so that $RG610 = R$ for a star with $R-I = 0$).
The $RG610$ magnitudes of the SN in the search observations were established
via comparison with several local field stars.  Finally, the SN $RG610$
magnitudes were $K$-corrected to ANDICAM $R$, using $K$-corrections
calculated from the measured transmission curves for the ANDICAM $R$
and Palomar-QUEST $RG610$ applied to SNIFS spectra on either side of maximum
light.  The process has an estimated error floor of about 0.05 mag.
A similar procedure was followed for the SNIFS $V$-band photometry.
The $RG610$ and SNIFS $V$ photometry have been merged, respectively,
with the ANDICAM $R$ and $V$ lightcurve points in the analysis to follow.

Finally, \citet{cbet2007if} and Yuan et al. (2010, in preparation) present
unfitered photometry obtained with ROTSE-IIIb that is valuable in
constraining the rise time (see Section \ref{subsec:arnett}).

% ============================================================================

\subsection{Host galaxy observations}
\label{sec:host}

\begin{figure}
\center
\resizebox{\columnwidth}{!}{\includegraphics{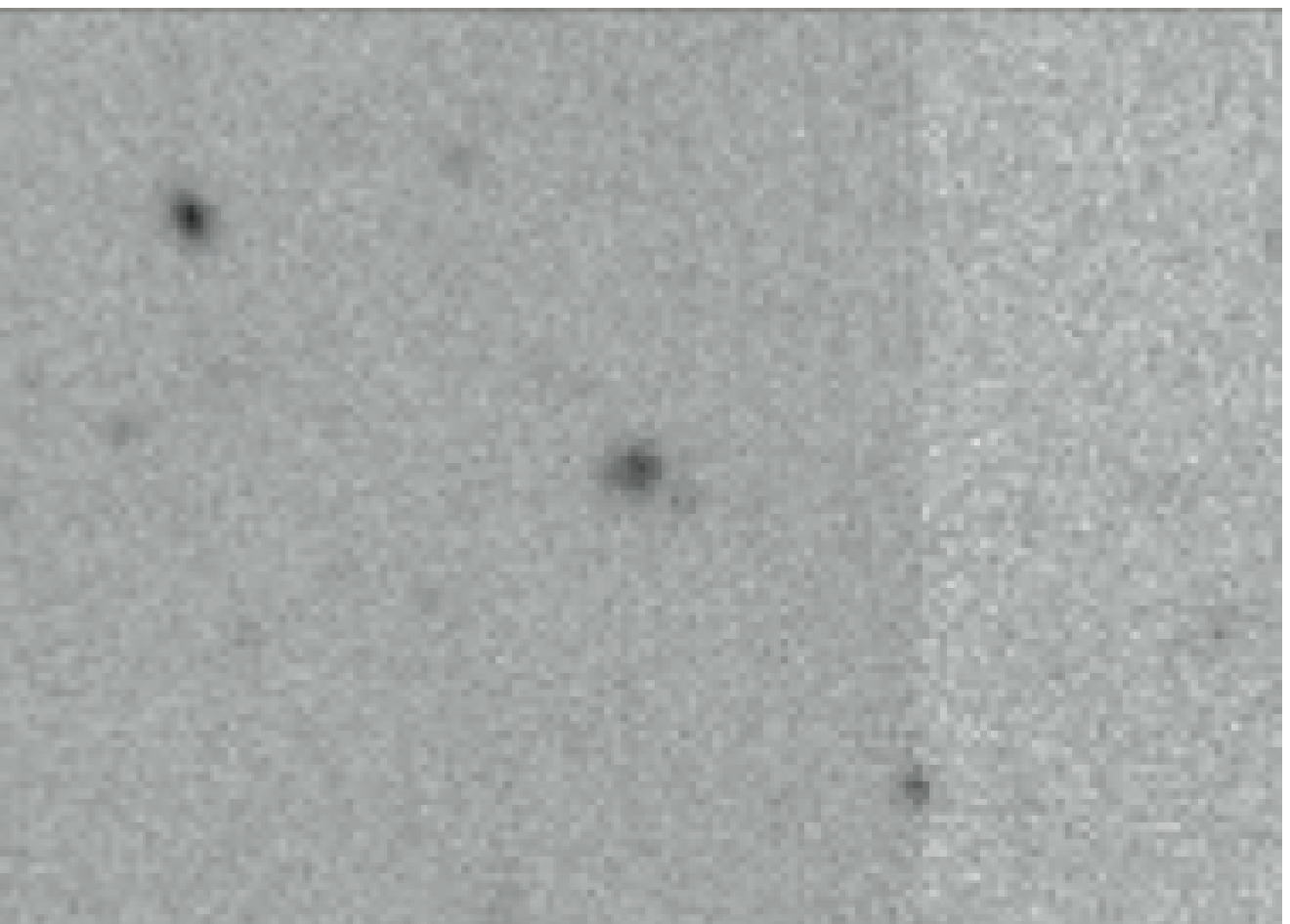}}
\caption{\small Image of the host galaxy of SN~2007if as co-added
from Keck LRIS imaging (north is up, east is left).
The FOV is 43\arcsec$\times$30\arcsec.}
\label{fig:host-thumb}
\end{figure}

The host of SN~2007if is not visible in pre-explosion survey images, and no
host spectroscopic features were detected in SNIFS spectra obtained while the
SN was being actively followed.  On 2009~August~24.6 UTC, imaging obtained
with the Low Resolution Imaging Spectrograph \citep[LRIS;][]{oke95}
on the Keck-I 10~m telescope revealed an extended source coincidence
with the original SN location.  Five exposures each of 100~s duration
were obtained using the LRIS blue-side camera equipped
with a g-band filter. The exposures were dithered to allow rejection of
cosmetic defects. The raw images were overscan-subtracted, flat-fielded
using dome flats, astrometrically calibrated using {\tt WCSTools}
\citep{mink06} and then coadded using {\tt SWarp} \citep{bertin02}. The
final image, shown in Figure~\ref{fig:host-thumb},
was segmented using {\tt SExtractor} \citep{bertin96} and
then flux calibrated using stars overlapping with SDSS-I, matched using
{\tt WESIX}\footnote{\url{http://nvogre.astro.washington.edu:8080/wesix/}}.
The resulting host location is
$\alpha = $~01:10:51.412,
$\delta = $~15:27:39.57
and the magnitude based on {\tt SExtractor mag\_iso\_cor}
after correcting for Galactic extinction of $A_g = 0.34$ is $g =
22.89\pm0.04$.  The host major and minor axes are $1 \farcs 5$ and
$1 \farcs 2$, respectively, uncorrected for seeing of $0 \farcs 76$ FWHM.

%%  g-Petrosian to mag_isocor using N stars with 17.5 < g < 22.0
%%  robust linear fit:   33.1165      + 0.0115909*(g-21)
%%                    +/- 0.0218525 +/- 0.0227103
%%  median:              33.1221
%%
%%  host mag_isocor = -9.89 (+/- 0.03) + 33.12 (+/- 0.02) = 23.23 +/- 0.04

Upon detection of the host, LRIS was reconfigured for spectroscopic
observations intended to measure the host redshift. On the blue side the
600 l~mm$^{-1}$ grism blazed at 4000~\AA\ was employed, covering 3500--5600~\AA,
while on the red side the 900 l~mm$^{-1}$ grating blazed at 5500~\AA\ was used
to cover 5440--7640~\AA. The D5600 dichroic was used to direct light to
the appropriate channel. A slit width of 1\arcsec\ was used, giving
resolutions of $\lambda/\Delta\lambda \sim 1200$ and $\sim 1650$ for the
blue and red sides, respectively. Four exposures of 900~s each were
obtained. These were overscan-subtracted, flat-fielded using internal quartz
flats, cosmic-ray rejected using \texttt{LAcospec} \citep{pvd01}, and then
wavelength-calibrated using arc and night sky lines. The reduced spectrum
revealed weak emission lines from H$\alpha$ and \ion{O}{2} $\lambda 3727$. The
resulting heliocentric redshift is $z=0.07416\pm0.00082$, and the resulting host
luminosity is $M_g=-14.10\pm0.07$ after correction for Galactic extinction.

%The low luminosity of the SN~2007if host strengthens the
%association of 2003fg-like SNe~Ia with young stellar populations in
%low-metallicity environments; we will publish our metallicity estimate in
%a separate paper \citep{childress}.  Such faint galaxies were not regularly
%searched prior to the advent of wide-area, untargeted supernova searches
%such as the Palomar-QUEST/SNfactory and ROTSE-III/Texas searches.
%Future wide-area searches will continue to tighten the constraints on the
%relative rates of these SNe and the luminosity function of their host
%galaxies \citep{young08}.

% ############################################################################

\section{Discussion}

Our extensive photometric and spectroscopic dataset allows us to examine
several unique features of SN~2007if. We begin with comparison to the
spectrum of the prototype --- SN~2003fg --- as well as of SN~2006gz,
focusing in particular on the evidence for unburned carbon in such
systems.  The velocity evolution is examined, especially in light of
the low velocities previously found for SN~2003fg and SN~2009dc. We then
examine the photometric properties and the influence of dust extinction,
in preparation for determining the mass of the SN~2007if progenitor.

% ============================================================================

\subsection{Spectral features and velocity evolution}

\begin{figure*}
\center
\resizebox{!}{8in}{\includegraphics{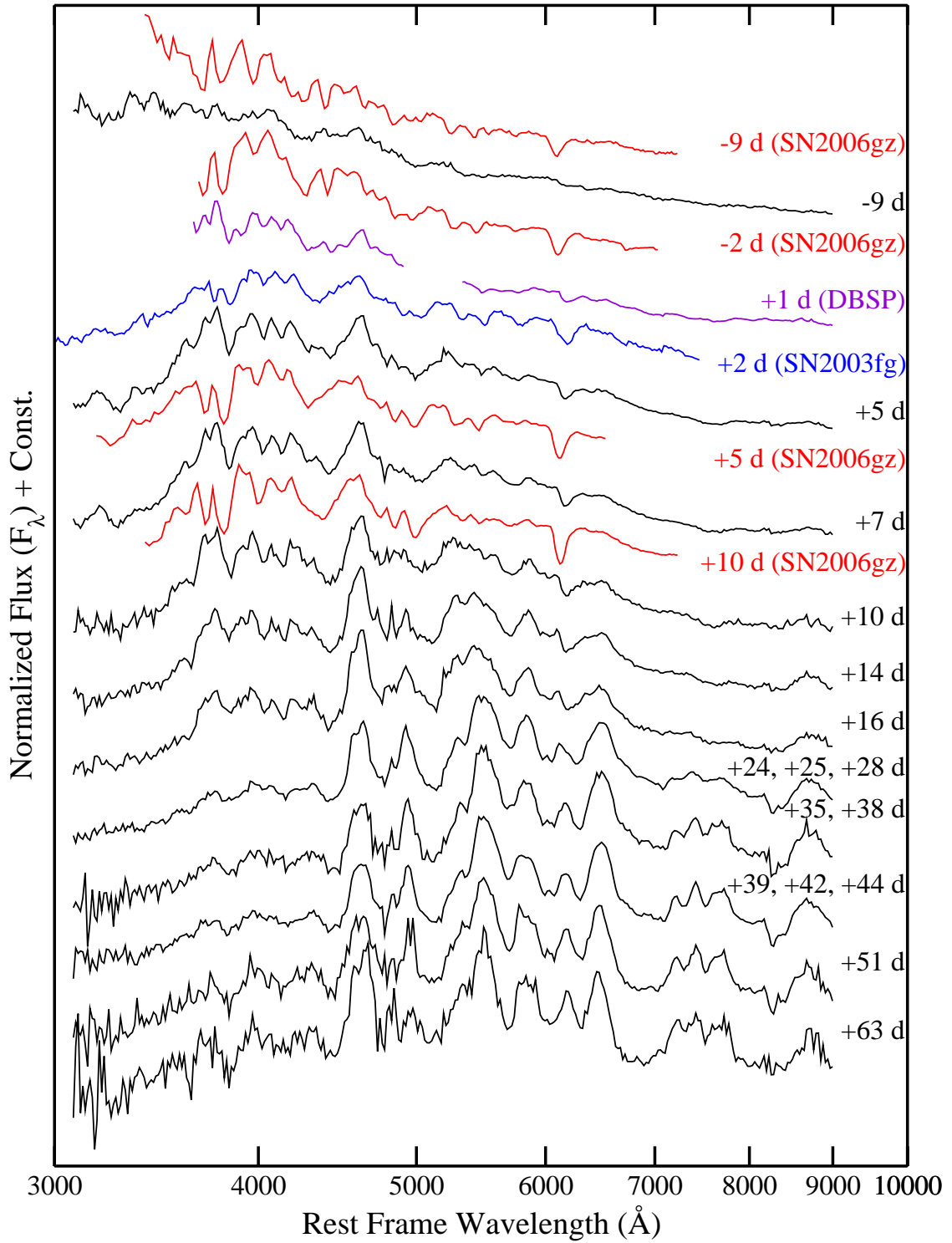}}
\caption{\small Selected rest-frame SNIFS spectra of SN~2007if
(black curves), interleaved with observations of SNe~2006gz
\citep[red,][]{hicken07} and 2003fg \citep[blue, the prototypical
super-Chandrasekhar SN~Ia candidate,][]{howell06}.  A spectrum obtained
just after maximum with the Double Spectrograph (DBSP, violet) on the
5.1~m Palomar Hale telescope is also included.  Spectra of SN~2006gz are
selected based on proximity in rest-frame phase for comparison.  The
$-9$~d spectrum of SN~2006gz has more developed absorption features than
SN~2007if at the same phase, making SN~2007if more difficult to
recognize as a SN~Ia.  SN~2007if is in some respects a better match to
SN~2003fg than SN~2006gz, but not in every case.  Note especially the
more blueshifted absorptions of SN~2006gz (especially \ion{Si}{2} around
6100~\AA).  All spectra shown have been binned to 1000~km~s$^{-1}$ and later
phase SNIFS spectra combined for presentation purposes.}
\label{fig:spec-comp}
\end{figure*}

Figure~\ref{fig:spec-comp} displays selected SNIFS spectra of SN~2007if,
along with selected published spectra from SN~2006gz and the only
spectrum of SN~2003fg available.  The SNIFS spectra of SN~2007if are
supplemented by the $+1$~d spectrum obtained at the Palomar Hale 5.1-m
telescope with the Double Spectrograph.  The spectral time-series of
SN~2006gz published by \citet{hicken07} extends from $-14$~d to $+11$~d
with respect to maximum --- the earliest spectrum from that event had
more pronounced absorption features clearly identifying it as a SN~Ia,
and also a prominent \ion{C}{2} $\lambda 6580$ absorption initially
attributed to a low velocity \ion{Si}{2} component \citep{prieto06}.  In
contrast, the earliest SNIFS spectrum of SN~2007if is relatively
featureless, and without prior knowledge of the host redshift, not
obviously from a SN~Ia.  In general, the absorption features common to
both SN~2006gz and SN~2007if appear to be broader, stronger, and more
highly blue-shifted in the former than in the latter --- note the faster
\ion{Si}{2} absorption around 6100--6200~\AA\ in SN~2006gz.  Around
maximum, the spectrum of SN~2007if seems to more closely resemble that
of SN~2003fg than it does SN~2006gz, in particular at the double-notch
absorptions around 4000--4200~\AA, but also note the double absorption
at 3700--3900~\AA\ missing from SN~2007if.  Comparison at much later
phases to SN~2006gz is not possible since that time-series ends much
earlier than that from SNIFS.  The detailed evolution at these phases is
rather slow, but these spectra appear quite similar to those of a normal
SN~Ia except for the lack of emission around 4000~\AA.

\begin{figure}
\center
\resizebox{\columnwidth}{!}{\includegraphics{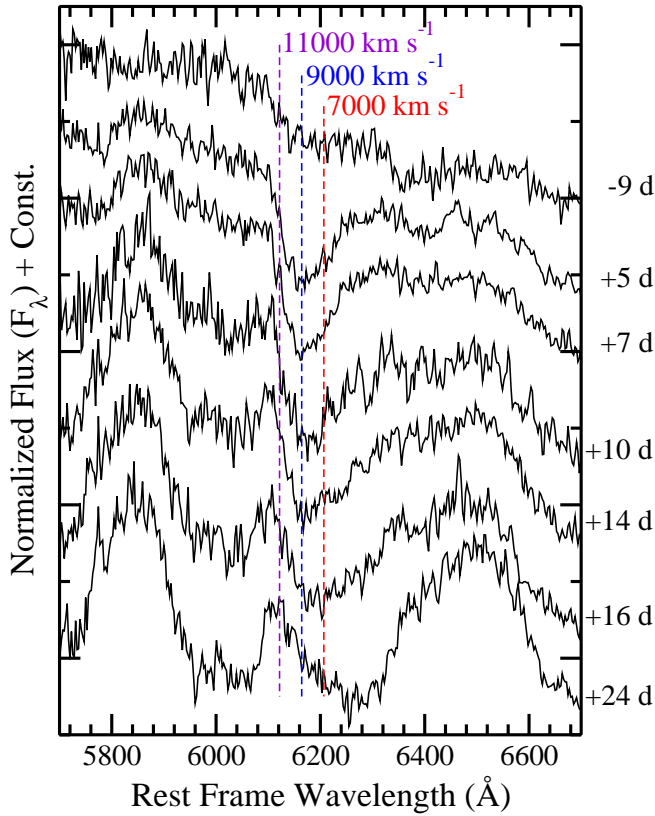}}
\caption{\small Detailed evolution of SN~2007if spectra around the
\ion{Si}{2} $\lambda 6355$ feature.}
\label{fig:vsi}
\end{figure}

In Figure~\ref{fig:vsi} we present the detailed evolution of the SNIFS
spectra in the region of the \ion{Si}{2} $\lambda 6355$ feature.  In the
earliest spectrum, the feature with the blue absorption edge at 6360~\AA\
may be a sign of \ion{C}{2} $\lambda 6580$
at a velocity of $\sim 10,000$~km~s$^{-1}$ --- suggesting the presence
of unburned carbon as in SNe~2006gz and 2009dc, though the line is much
weaker.  Due to the redshift of SN~2007if, this region overlaps telluric
$B$-band absorption, but we have confirmed the accuracy of our telluric
correction and believe the notch to be a real SN feature.
The evidence for \ion{C}{2} at this wavelength in SN~2003fg is more
circumstantial owing to that spectrum's lower S/N at these wavelengths
and the putative identification of \ion{C}{2} $\lambda 4267$ (but see
below).  Immediately blueward is the expected position of \ion{Si}{2}
$\lambda 6355$, at which an absorption ``slump'' is detectable.  The
sloping continuum and proximity to \ion{C}{2} make the line profile
difficult to measure directly; we factor out the continuum in our
measurements of the absorption minimum described below, and include related
uncertainties in our error bars.  In any case, the shallow inflection
near 9,000~km~s$^{-1}$ and the blue edge near 11,000~km~s$^{-1}$ (both
relative to 6355~\AA), coincident with the edge of the \ion{Si}{2}
feature in the post-maximum spectra, suggest that the photosphere may
have receded to the post-maximum velocity at a phase of $-9$~d.  The
relative robustness of this edge across phase may suggest that the layer
of Si-peak elements production extends only to 11,000~km~s$^{-1}$.

\begin{figure}
\center
\resizebox{\columnwidth}{!}{\includegraphics[clip=true]{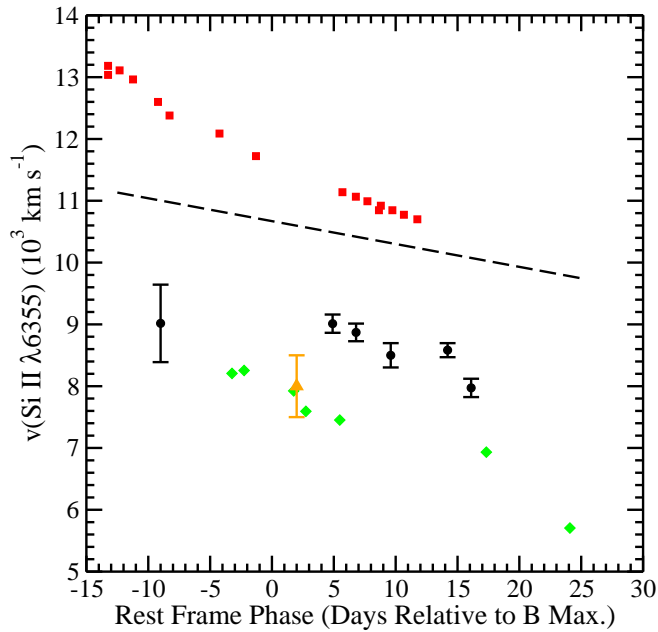}}
\caption{\small Comparison of the evolution of the \ion{Si}{2}
absorption feature in SN~2007if (black circles) with SNe~2006gz (red
squares), 2009dc (green diamonds) and 2003fg (orange triangle).  The
mean evolution of \ion{Si}{2} velocity in normal ``LVG'' subclass SNe~Ia
\citep{benetti05} is presented as the dashed line.}
\label{fig:vsi-comp}
\end{figure}

Figure~\ref{fig:vsi-comp} compares the evolution of the position of
the \ion{Si}{2} $\lambda 6355$ absorption minimum in SNe~2007if,
2006gz, 2009dc, and 2003fg.  At least in the case of SN~2007if, by the
time of the next SNIFS observation after day $+16$ (on day $+24$), the
SN spectrum has evolved too much to associate \ion{Si}{2} $\lambda
6355$ with any given spectral feature without peril.  We exclude the
DBSP spectrum from this analysis, since it was obtained at lower
resolution and in poor conditions; however a measurement from the
profile is consistent with the SNIFS observations a few days
later.  The measurements of \ion{Si}{2} $\lambda 6355$ on SNIFS spectra
were made as follows:  bins in each spectrum immediately
to the right and left of the line feature were used to fit a linear
local continuum, $F_{\lambda,\mathrm{cont}} = a + b\lambda$.
The spectrum in the region of the line was then fitted using
$F_\lambda = F_{\lambda,\mathrm{cont}} \times \phi(\lambda)$ where
the absorption line was modeled as a skewed Gaussian,
\begin{equation}
\phi(\lambda) = 1 - c [1+d(\lambda-\lambda_0)^3] \,
           e^{-(\lambda-\lambda_0)^2/2 \sigma^2}
\end{equation}
The line velocity was determined by solving
$d\phi/d\lambda = 0$ via the Newton-Raphson method.
The error bars on the procedure were determined through a bootstrap
Monte Carlo in two stages.  In the first stage, values of $a$ and $b$
representing possible continua $F_{\lambda,\mathrm{cont}}$ were sampled using
the covariance matrix of the local continuum fit;
for each candidate continuum, the line profile $\phi(\lambda)$ was then refit
holding $a$ and $b$ fixed, and values of
$c$, $d$, $\lambda_0$ and $\sigma$ were sampled using the covariance matrix
of the fit to $\phi(\lambda)$.  The final velocity values and their errors
were determined as the mean and standard deviation of the distribution of
absorption minimum velocities thus generated.
%% [This was the Savitzky-Golay approach text]
%% The measurements of \ion{Si}{2} $\lambda 6355$ on SNIFS spectra
%% were made using a Savitzky-Golay (SG) smoothing applied to a
%% restricted range of the spectra (5800-6400~\AA). To estimate the
%% velocity, the minimum of the SG regularized spectrum will give us the
%% shifted wavelength searched within the 6000-6210~{\AA} window. The SG
%% regularization use a second order polynomial fit inside a running
%% window whose width was set to minimize the global function
%% approximation error on the spectral range defined above. This width is
%% optimized independently for each spectrum, taking into account the
%% variance of each bin and the bin-to-bin correlation. The statistical
%% error on each measurement is evaluated from the velocity
%% distributions of 1000 simulated spectra with the same signal to noise
%% and bin-to-bin correlation as the initial one. A systematic error is
%% added quadratically, it was estimated by changing the optimal window
%% size $\pm$ 15\%, which reflects the uncertainties in the optimal size
%% coming from uncertainties in the weight matrix.
We can see in Figure~\ref{fig:vsi-comp} the measurements of the velocity and
their error with this method for each spectrum of the study. Clearly, the
average measured characteristic ejecta velocities of SNe~2007if and 2009dc are
much more consistent with that measured for SN~2003fg, the prototypical
super-Chandrasekhar event.  The shallow slope of the velocity evolution
($\dot{v} = 34 \pm 15~\mathrm{km~s^{-1}~day^{-1}}$) is characteristic of the
low-velocity-gradient \cite[LVG;][]{benetti05} subclass.

To further characterize the SN ejecta, we have analyzed the $+5$ day SNIFS
spectrum using the automated
SN spectroscopic direct analysis program, \texttt{synapps} \citep{rct09}.  This code incorporates
the familiar SYNOW-style \citep{jb90,bbj03} parameterized model as part
of an objective function in a multidimensional nonlinear optimization
problem.  From a good fit the presence of given ions may be ruled out,
and for ions that are positively identified, corresponding intervals of
ejecta velocity are constrained.  Such a parameterized approach is
indispensable in the analysis of such particularly unusual events as
SN~2007if for which no reasonable detailed theoretical prediction
otherwise exists, and will continue to play a vital role as long as new
types of transients are discovered.

\begin{figure}
\center
\resizebox{\columnwidth}{!}{\includegraphics[clip=true,bb=60 40 730 540]{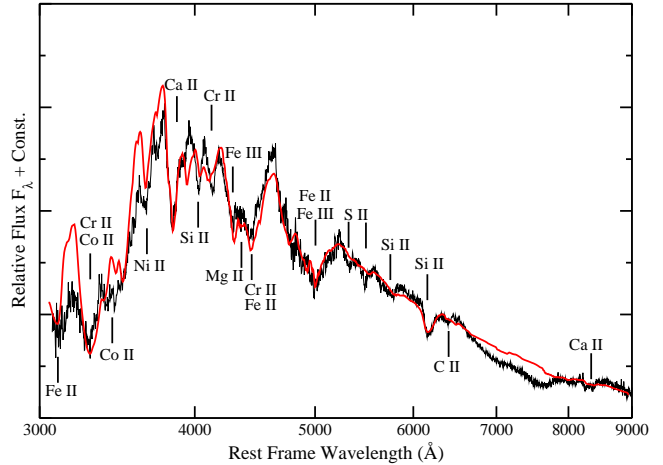}}
\caption{\small SYNOW fit to the SNIFS day $+5$ spectrum.}
\label{fig:synow}
\end{figure}

A fit to the day $+5$ spectrum appears in Figure~\ref{fig:synow}.  The
entire SNIFS wavelength range is included in the fit.  Familiar SN~Ia
ion species \ion{Mg}{2}, \ion{Si}{2}, \ion{S}{2}, \ion{Ca}{2},
\ion{Fe}{2} and \ion{Fe}{3} produce unambiguous spectroscopic absorption
signatures.  Blanketing by iron-peak elements is invoked to suppress UV
flux, but the optimized model may overcompensate for the tendency of the
Sobolev approximation to underblanket here. The \ion{Si}{2} and
\ion{S}{2} features are rather weak, a trait in common with the so-called
overluminous or SN~1991T-like SNe~Ia.  The \ion{Ca}{2} H\&K absorption
at 3800~\AA\ is extremely sharp and weak compared to normal events, as
was noted in the case of SN~2003fg.  The highly blended regions centered
on 4500~\AA\ and 5000~\AA\ are actually dominated by \ion{Fe}{3},
again harkening to SN~1991T-like events.  The relatively weak IME
signatures and the prevalence of \ion{Fe}{3} indicate high temperatures
in the ejecta; in SN~1991T this has been attributed to the presence of
\nickel\ in the outer layers of the ejecta \citep{mazzali95}.
We also detect a weak signature of \ion{C}{2} at around $8000$~km~s$^{-1}$,
i.e., at velocities comparable to the \ion{Si}{2} feature.

Of particular interest is the pair of sharp notches at 4030~\AA\ and
4130~\AA.  In \cite{howell06}, the bluer notch is attributed to
\ion{Si}{2}, and the redder to \ion{C}{2}.  In the fit presented in that
work, a strong \ion{C}{2} $\lambda 6580$ absorption was predicted, but
the mismatch between the prediction and observations was attributed
mainly to the poor S/N in those spectra.  The S/N in the SNIFS spectrum
is much higher, but the \ion{C}{2} $\lambda 6580$ is very weak.  While
NLTE effects may be in play \citep{rct07}, we considered further
alternative explanations for the 4130~\AA\ feature.  An interesting
possibility is \ion{Cr}{2} absorption --- the inclusion of this ion also
helps produce the strong emission at 4600~\AA, and contributes to the
absorption at 3280~\AA, along with other iron-peak elements.
\ion{Cr}{2} is not without precedent in SN~Ia spectra, though its appearance
in more normal events seems limited to postmaximum epochs \citep{branch08}.
However, detailed reconstruction of the 4000--4130~\AA\ region remains
difficult --- in particular, \ion{Ni}{2} is invoked to explain some of
the UV line blanketing, but produces an extra notch at 3950~\AA.

\subsection{Maximum-light behavior, colors and extinction}
\label{sec:lc-bvri}
%\subsection{Rest-frame $BVRIJ$ light curve}
%\label{subsec:bvrij}

\begin{figure}
\center
\resizebox{\columnwidth}{!}{\includegraphics{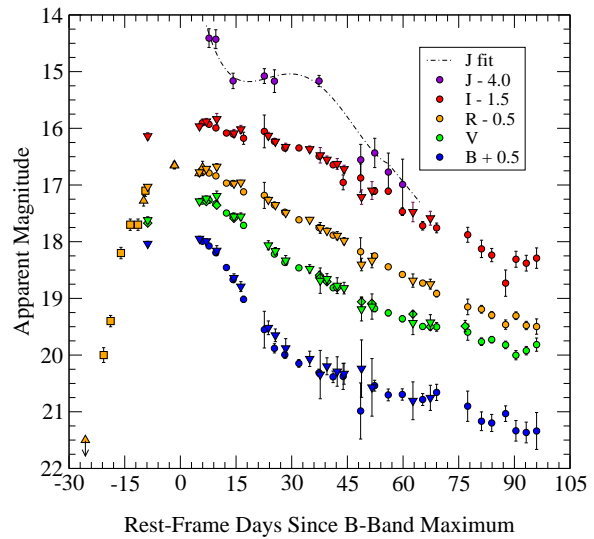}}
\caption{\small Rest-frame Bessell $BVRIJ$ light curve for SN~2007if
from ANDICAM+SNIFS+QUEST.  Upright triangles:  QUEST $RG610$,
$K$-corrected to rest-frame Bessell $R$.  Diamonds:  SNIFS $V$ from
photometric imaging channel, $K$-corrected to rest-frame Bessell $V$.
Inverted triangles:  Rest-frame Bessell $BVRI$ magnitudes synthesized
from SNIFS flux-calibrated spectroscopy.
Circles:  ANDICAM $BVRIJ$, $K$-corrected to the respective rest-frame
Bessell filters.
Squares:  ROTSE-III points from \cite{cbet2007if}.
A fourth-order polynomial fit to the $J$-band data is also shown.}
\label{fig:lc-bvrij}
\end{figure}

\begin{figure}
\center
\resizebox{\columnwidth}{!}{\includegraphics{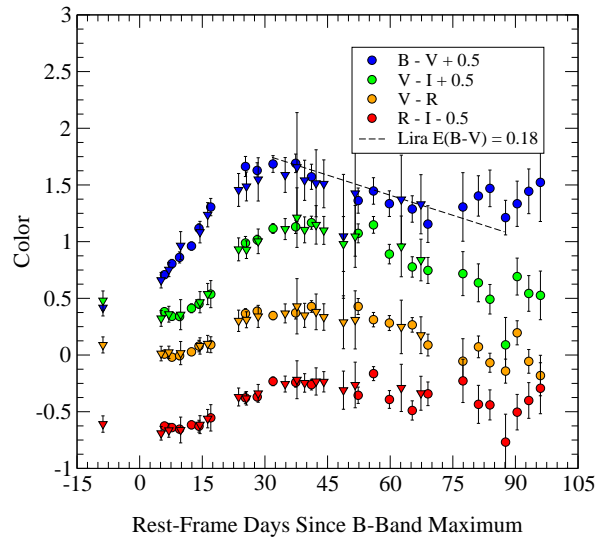}}
\caption{\small Rest-frame Bessell color evolution for SN~2007if
from ANDICAM (circles) and SNIFS spectrophotometry (inverted triangles).
A fit to the Lira relation with $E(B-V) = 0.18$ is shown.}
\label{fig:lc-bvlira}
\end{figure}

The rest-frame Bessell $BVRIJ$ light curve of SN~2007if is given in
Table~\ref{tbl:lc-rest} and in Figure~\ref{fig:lc-bvrij}.
The color evolution is shown in Figure~\ref{fig:lc-bvlira}.
An accurate estimate of the
luminosity and \nickel\ mass for SN~2007if requires some care, due in part to the sparse multi-band
light curve coverage around maximum light.  
In this section we explore two
different methods of interpolating the light curves to maximum, derive
the observed color of the SN, and comment briefly on reddening by dust in
the host galaxy.

Fitting the multi-band light curve of SN~2007if with existing SN~Ia templates
trained on normal SNe~Ia is of course problematic.
As an example, the SALT2 model \citep{salt2} fit to all four bands is poor
($\chi^2_\nu = 930.3/122 = 7.6$), above the threshold used by the SNLS to
photometrically screen events as ``probable'' SNe~Ia \citep{sullivan06}.
The event would therefore have been deprioritized for spectroscopic follow-up
by SNLS and might not have been studied further.  The fit is especially poor
in $R$ and $I$, which lack the distinct second maximum typical of normal
SNe~Ia; such behavior also was noted in SN~2006gz by \cite{hicken07}
and in SN~2009dc by \cite{yamanaka09}.

We can nevertheless attempt to use SALT2 simply as a method for
interpolating the shape of the light curve around maximum light, taking
the $\chi^2$ and fitted parameters $x_1$ and $c$ provisionally.
To minimize the impact of details of the SALT2 spectral model on the outcome,
we use SALT2 in the rest frame, include SALT2 light curve model errors in
the fitting, and we fit $BVR$ bands only, excluding the clearly
discrepant $I$.  This procedure gives a more reasonable
$\chi^2/\mathrm{\nu} = 144.8/93 = 1.6$,
with a date of $B$-band maximum around MJD~54348.4 (2007~September~6),
a stretch of $1.15$ ($x1 = 1.83$, consistent with SN~2003fg)
and a very red color ($c = 0.24$).
The best-fit magnitudes at $B$-band maximum are
$m_B = 17.34 \, \pm \, 0.04$, % 17.29 +/- 0.05 for BV fit only
$m_V = 17.18 \, \pm \, 0.05$, % 17.17 +/- 0.05 for BV fit only
$m_R = 17.14 \, \pm \, 0.06$  % 17.14 +/- 0.06 for BV fit only
(statistical + $K$-correction errors added in quadrature), giving $B-V = 0.16$.
% RS:  K-correction errors are everywhere less than 0.02 mag
The value of $m_R$ so derived is also consistent with the observed unfiltered
magnitude at maximum \citep[calibrated to USNO-B~1.0~$R$][]{usno}
observed by ROTSE-IIIb
\citep{cbet2007if} and the QUEST-II $RG610$ point at that date.

As a cross-check, we fit a cubic polynomial to the rest-frame $BVR$ points
before MJD~54374.0 (three weeks after the SALT2 date of $B$-band maximum).
This gives maximum magnitudes $m_B = 17.33 \pm 0.03$, $m_V = 17.25 \pm 0.03$,
and $m_R = 17.19 \pm 0.02$, with the date of $V$-band maximum consistent with
the SALT2 fits to within $\pm2$~days.  The uncertainties on the date of
maximum are larger with this approach, since each band can vary independently;
however, the results are roughly consistent with SALT2.  We take the results
of the SALT2 fit as our fiducial values in further discussion ---
in particular the phase of maximum light at MJD~54348.4, in which case
the SMARTS observations begin at day $+5$ after maximum light.

% RS:  JD of max from SNIFS is 2454347.1; it's 2454348.7 for my fit,
% or an offset of -1.6 days.  That brings the first spectrum to -10 days.
% Need to fix all this in the figure.  Remember that MJD = JD - 2400000.5,
% so that last 0.5 makes a slight difference.

Both of the above extrapolations imply that SN~2007if is redder than
the typical SN~Ia at maximum light.  To assess the impact of dust extinction,
we use the Lira relation \citep{phillips99,csp09} and the equivalent width of
\ion{Na}{1}~D absorption \citep{tbc02} (``TBC relation'').  Two possible
slopes for the TBC relation are given, one shallow
($E(B-V) \sim 0.5 \times EW$(\ion{Na}{1}~D)) and one steep
($E(B-V) \sim 0.15 \times EW$(\ion{Na}{1}~D)).

The $B-V$ color evolution of SN~2007if (see Figure~\ref{fig:lc-bvlira})
has a slope compatible with the the Lira relation \citep{phillips99}
with a fitted color excess of $E(B-V)_\mathrm{host} =
0.18 \pm 0.04 \ \mathrm{(stat)} \ \pm 0.06 \ \mathrm{(sys)}$.
Assuming the Lira relation holds
(which for this unusual SN is by no means clear),
the intrinsic color of the SN is $B-V = -0.02$.

\newcommand{\ewna}{\ensuremath{EW}(Na~I~D)}

A noise-weighted co-add of the SNIFS spectra reveals no \ion{Na}{1}~D
absorption at the redshift of the host.  To derive an upper limit on
\ewna\ while allowing for uncertainty in the redshift,
we construct the probability surface over equivalent width and redshift
for a joint fit of the \ion{Na}{1}~D line profile within the appropriate
wavelength range for all observed spectra of SN~2007if.  In the fit,
the local continuum is estimated separately for each spectrum as a cubic
polynomial, and the absorption is fully modeled using separate wavelengths
for both \ion{Na}{1}~D lines using the SNIFS instrumental resolution
of 6~\AA.  The resulting probability density was then marginalized over
the redshift error to obtain the final estimate.  For the Milky Way dust
absorption, we derive \ewna~$= 0.51^{+0.04}_{-0.05}~\AA$,
which under the ``shallow'' TBC relation corresponds to
$E(B-V) = 0.072$, in good agreement with the reddening from \cite{sfd}.
No absorption is detected from the host, and we derive an
upper limit \ewna~$< 0.14~\AA$ (95\% CL),
which corresponds to $E(B-V) < 0.032$ even under the ``steep'' TBC relation.

A similar discrepancy between the Lira relation and \ion{Na}{1}~D absorption
was noted by \cite{yamanaka09} for SN~2009dc; their work ended up adopting
the estimate from \ewna.  Under the same assumption for SN~2007if,
host reddening is therefore unlikely to account for a significant share
of SN~2007if's deviation from the colors of normal SNe~Ia.
The color difference is probably real and intrinsic.
If the entire color difference were due to reddening with $R_V \sim 3.1$,
this would make SN~2007if nearly a full magnitude brighter at maximum
than SN~2003fg.  However, given the disagreement between the two host
reddening estimates, we revisit the potential impact of uncontrolled host
reddening on our results in Section \ref{sec:reddening}.
\emph{The primary analysis to follow in our paper assumes zero host galaxy
reddening.}

% ----------------------------------------------------------------------------
% 95% CL upper limit on V-band extinction from coadd?
% TBC law:  E(B-V) = -0.04 + 0.51*EW(Na I) or -0.01 + 0.16*EW(Na I)
% My estimate:  if unresolved -- say 3 AA resolution at 6000 AA, or 150 km/s
% Line depth will be suppressed by a factor of q = 1/sqrt(1+(3 AA/w)^2)
%    -- RS:  The above demonstrates that the upper limit is extremely
%       sensitive to the assumed velocity scale of the host!
%       For dinky host the line is really underresolved and there's
%       no way SNIFS is going to see it.
% Say w = 100 km/s = 2 AA, then q = 0.5 -- the limit degrades by this much
% Now suppose S/N = +/- 1/(0.000003125/0.00010) = 32, let's say 30, near peak
% so let's say we have 10 spectra with S/N ~ 20 or so.
% Co-add 10 spectra gives S/N ~ 60, or EW < (1/60)*3 AA = 0.05 AA (1 sigma)
% This means EW(Na I) < 2/0.5 * 0.05 AA = 0.2 AA at 95% CL (roughly),
% meaning E(B-V) < 0.1 from the host galaxy at 95% CL.
% ----------------------------------------------------------------------------
% Lira relation (E(B-V) = 0): B-V = 0.725 - 0.0118*((t-tVmax)-60)
% with intrinsic dispersion about 0.05 mag from SN to SN.
% From our fit:  E(B-V) = 0.18 +/- 0.04 (stat) +/- 0.05 (sys)
% Suggests B-V = 0 if it's all reddening.
% This would mean M_V = -20.72 even with RV = -2.1 if it's all intrinsic.
% Most likely Lira is an overestimate.
% ----------------------------------------------------------------------------

Adopting the magnitudes at maximum light from the SALT2 fits to the
rest-frame light curve and a CMB-frame redshift of
$z_\mathrm{CMB} = 0.0731\pm.0013$ 
(including a 300~km~s$^{-1}$ peculiar velocity)
gives a Hubble-flow distance modulus of
$\mu = 37.57\pm0.03$ for the concordance cosmology
\citep[$h = 0.71$, $\Omega_M = 0.27, \Omega_\Lambda = 0.73$; see][]{nedwright}
implies absolute magnitudes $M_B = -20.23$, $M_V = -20.39$ assuming zero
reddening from the host galaxy.  This makes SN~2007if the brightest
super-Chandrasekhar-mass SN~Ia candidate yet discovered.
The SN is at least some 1.3 mag brighter in $V$ than a ``normal'' $x_1 = 0$
SN~Ia of the same color, and too bright for its SALT2 value of $x_1$ by about
1.1 mag
\citep[taking $M_V(x_1=0,c=0) = -19.07$, $\alpha = 0.13$, $\beta = 1.77$,
after][]{salt2}.

% RS:  Worth noting that the full SALT2 fit from the Louisiana run of SNIFS
% spectrophotometry gets a somewhat less extreme red color, c = 0.16.
% To the extent that we believe that, it really is red.
% The SNIFS fit suggests the SN should be brighter at maximum, m_B = 17.24.
% The magnitudes from the ``normal'' SALT2 fit to the ANDICAM data, using the
% SALT2 K-corrections, are more like m_B = 17.35, m_V = 17.18, with a color
% of c = 0.23.

The temporal sampling and S/N in ANDICAM $J$ are much lower than
in the optical bands, due to the limited time available in our SMARTS
observing program.  However,
the light curve suggests a second maximum starting 15--20~days after $B$-band
maximum light, lasting until about 40~days after $B$-band maximum light and
then dropping dramatically.  The height of the first maximum is not clear from
the data; the second maximum probably occurs 28--34~days after maximum light,
at a height of $J \sim 19$.  Figure~\ref{fig:lc-bvrij} shows a fit of a
fourth-order polynomial to the data within the range of phase coverage,
showing a possible position for the second maximum.

The estimate of $J$-band flux at maximum light is similarly problematic.
If the $J$-band light curve near peak is similar to the behavior e.g.
of $R$-band, extrapolated backwards nearly linearly without a pronounced
maximum, then maximum light might be only as bright as the brightest observed
point, or $J_\mathrm{max} = 18.4$.  If the first maximum is more pronounced,
as in the high-\nickel\ light curves of \cite{kasen06}, then it could be as
bright as $J_\mathrm{max} = 17.8$.  We take these estimates as representative
lower and upper limits.

Given these estimates of the flux in different bands at $B$-band maximum
light, and a spectrum similar to the spectrum at +5 days, we estimate
a maximum bolometric luminosity of
$(3.22 \pm 0.15) \times 10^{43}$ erg~s$^{-1}$.

% ============================================================================

\subsection{Light curve comparison to SN~2003fg}

\begin{figure}
\center
\resizebox{\columnwidth}{!}{\includegraphics{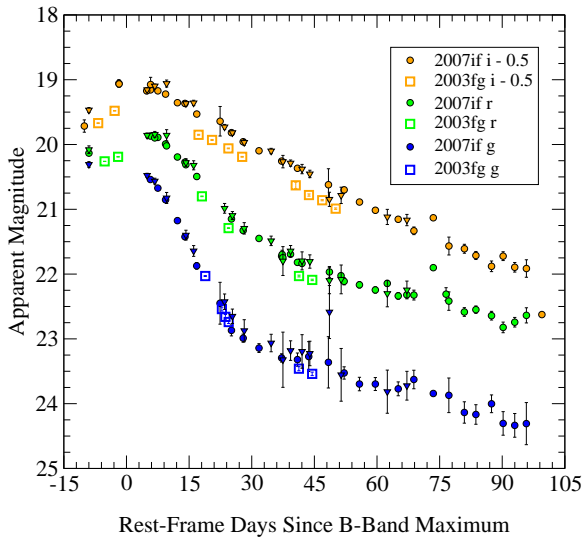}}
\caption{\small Direct comparison of SN~2007if multi-band light curves with
SN~2003fg in the latter's observer frame.  Solid circles:  ANDICAM+QUEST+SNIFS
$BVR$ for SN~2007if, $K$-corrected to SNLS $gri$ at $z = 0.244$.
Open squares:  SNLS $gri$ observer-frame ($z = 0.244$) light curves
for SN~2003fg \citep[see][]{howell06}.}
\label{fig:lc-snls}
\end{figure}

Figure~\ref{fig:lc-snls} shows a direct comparison between the light curves
of SN~2007if and SN~2003fg.  Since a spectral time series which might yield
accurate $K$-corrections for SN~2003fg is not available, we have
instead $K$-corrected our SN~2007if light curve to the observer-frame $gri$
bandpasses of SN~2003fg at $z = 0.244$.

We can see from this comparison that SN~2007if and SN~2003fg have very
similar decay behavior.  SN~2003fg is certainly fainter than SN~2007if
by about 0.3 mag.  The relative color and light-curve phase between the
two SNe is uncertain due to the lack of maximum-light coverage in $g$-band
for SN~2003fg.  SN~2003fg is probably bluer than SN~2007if, but not by more
than about 0.1~mag, or $B-V = 0.06$; this is difficult to reconcile with
the reported $B-V = -0.15$ \citep{howell06,hicken07}.
We shall return to the interpretation of possible color
differences between the two SNe below; it is unclear that the excess should
be interpreted in terms of dust extinction.

% ############################################################################

\section{Synthesized nickel mass in SN~2007if}

The maximum magnitude $R = 17.2$ observed in the search and follow-up images
suggest a highly overluminous ($M_V \sim -20$) explosion for SN~2007if, similar
to SNe~2003fg and SN~2009dc.  This in turn suggests a very large mass of
\nickel\ --- could it be in excess of \mch?  
In this section we calculate a bolometric light curve of the SN,
and from an estimate of the bolometric luminosity at maximum light, derive the
\nickel\ mass using Arnett's rule \citep{arnett82}.

% ============================================================================

\subsection{Bolometric light curve}
\label{sec:bolo-lc}

Bolometric light curves of SNe~Ia are useful for estimating the \nickel\ mass
\citep{arnett82}, from the bolometric luminosity at peak, and the total
ejected mass, from an estimate of the energy deposition in the ejecta from
decay of radioactive \nickel\ and \cobalt\ \citep{stritz06}.
\cite{maeda09b} characterize each SN by the decay time of the bolometric
light curve.

To produce a bolometric light curve for a given SN~Ia,
\cite{stritz06} use a parametrized model to interpolate the multi-band light
curve in time, which is then numerically integrated in wavelength to find
the total flux over bands as a function of time.  We choose to take a
slightly different approach using the observed spectra of SN~2007if,
similar to that used by \cite{howell09}.
For each quasi-simultaneous set of ANDICAM $BVRI$ observations, we deredden
and deredshift the SNIFS spectrum nearest in time, then multiply it by a
smooth function (in this case a cubic polynomial) fitted so that the synthetic
photometry from the resulting spectrum matches, in a least-squares sense,
the corresponding ANDICAM imaging photometry in each band.
We then calculate the bolometric flux as the integral of the SNIFS spectrum
over all rest-frame wavelengths from 3100--9000~\AA.  Occasionally
($< 10\%$ of the time) an ANDICAM observation in one band is missing
due, e.g., to S/N considerations from intermittent clouds or bright
moonlight; since the phase coverage is overall still exemplary, such gaps
are covered via linear interpolation between adjacent light curve points,
and a statistical error bar assigned to the interpolated value based on the
S/N of the adjacent points.  This never occurs in more than one
of the bands at a time.
As a cross-check on the effects of the evolution of spectral features on the
bolometric flux thus reconstructed, we evaluated each bolometric light curve
point using the two spectra bracketing it in time, and took the difference
between the measurements as an estimate of the systematic error.
This difference was found always to be less, and usually much less,
than the statistical error on each light curve point.

This handles only the optical section of the light curve, however.
Given the discussion in Section \ref{subsec:kasen-ir} below, we can expect much
of the flux to be reprocessed into the NIR.  To account for this, we repeat
the above procedure for the NIR, by normalizing the synthetic $J$ magnitudes
over the spectra of SN~1999ee to match the observed $J$-band photometry.
We expect systematic errors resulting from this approximation to be small,
since SN~1999ee is itself an overluminous LVG SN~Ia.
We then integrate the flux from 9700--24800\AA.
To compensate for the poor phase coverage and S/N in ANDICAM $J$,
we normalize to the fourth-order polynomial fit in Figure~\ref{fig:lc-bvrij}.
We estimate, or place limits on, the $J$-band contribution outside the bounds
of phase coverage with a piecewise linear extrapolation based on the $J$-band
light curves from \cite{kasen06}.  Most of the high-\nickel\ models show a
decline of 0.08~mag~day$^{-1}$ after the second peak, which we adopt for
rest-frame day $+55$ to day $+70$.  The subsequent behavior is uncertain and
we allow for the possibility of a decline anywhere in the range
0.02--0.06~mag~day$^{-1}$.
These limits and uncertainties are propagated into the bolometric light curve
error bars.  We find that the NIR correction to the bolometric luminosity
ranges from 30\% of the optical luminosity
(near the second maximum) to 10\%--15\% after day $+60$.  The NIR
correction is less than 5\% around $B$-band maximum light, so that the
expected systematic error on the \nickel\ mass is small.

\begin{figure}
\center
\resizebox{\columnwidth}{!}{\includegraphics[clip=true]{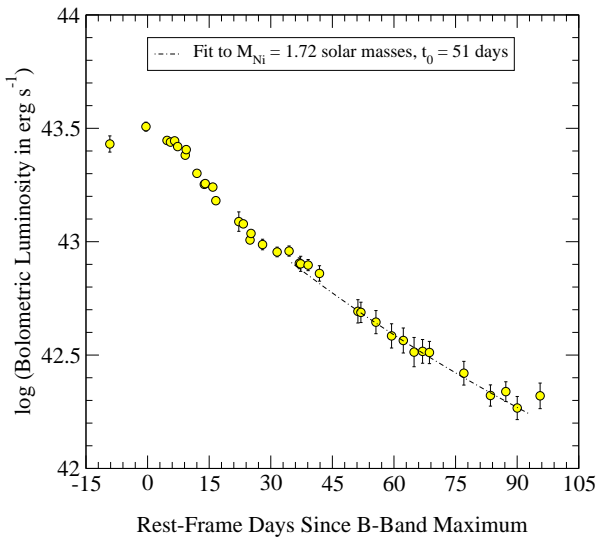}}
\caption{\small Rest-frame bolometric light curve of SN~2007if,
synthesized from SMARTS $BVRIJ$ data, SNIFS spectroscopy, and NIR spectra
of SN~1999ee taken with ISAAC on the VLT at Cerro Paranal \citep{hamuy02}.
A fit of the radioactive energy deposition curve to the data taken
more than 60~days after explosion is shown.}
\label{fig:lc-bolo}
\end{figure}

The resulting bolometric light curve is shown in Figure~\ref{fig:lc-bolo}.

% ============================================================================

\subsection{Estimate of the \nickel\ mass from Arnett's rule}
\label{subsec:arnett}

% RS:  We can do this by warping the SNIFS spectra to match the ANDICAM
% photometry, then integrating those numerically.  The SNIFS spectra thus
% serve as our wavelength interpolation in such cases.

To estimate the \nickel\ mass, we calculate the rate of radioactive energy
deposition in the ejecta,
\begin{eqnarray}
L_\mathrm{rad} & = & N_\nickel \lambda_\nickel Q_{\nickel,\gamma} \,
   \exp(-\lambda_\nickel t)
   \nonumber \\
& + & N_\nickel
   \frac{\lambda_\cobalt \lambda_\nickel}{\lambda_\nickel-\lambda_\cobalt}
      \left( Q_{\cobalt,e^+} + Q_{\cobalt,\gamma} \right) \nonumber \\
& \times & 
      \left( \exp (-\lambda_\cobalt t) - \exp (-\lambda_\nickel t) \right),
\label{eqn:opacity}
\end{eqnarray}
where $t$ is the time since explosion,
$N_\nickel = M_\nickel/(56~\mathrm{AMU})$
is the number of \nickel\ atoms produced in the explosion,
$\lambda_\nickel$ and $\lambda_\cobalt$ are the decay constants for
\nickel\ and \cobalt\ (lifetimes 8.8~days and 111.1~days) respectively,
and $Q_{\nickel,\gamma}$, $Q_{\cobalt,\gamma}$ and $Q_{\cobalt,e^+}$
are the energies released in the different stages of the decay chain
\citep{nad94}.  We adopt an efficiency factor $\alpha$ of order unity
such that $L_\mathrm{bol} = \alpha L_\mathrm{rad}$
\citep{hk96,jbb06}.  We use a fiducial value of $\alpha = 1.2$
\citep{jbb06,howell06,howell09}, within the range $0.8 < \alpha < 1.6$
representative of various models \citep{jbb06}.  Given the quality of our data,
the uncertainty on $\alpha$ is the limiting systematic error in our analysis.

Since ROTSE-IIIb detected the SN at unfiltered magnitude $19.5 \pm 0.1$
as early as August~16.3 UT \citep{cbet2007if}, the observer-frame
rise time must be at least 20~days.  Palomar-QUEST detects no object at the
SN~location on August~9.0, however, placing an upper limit of $RG610 > 21.5$
at 95\% confidence; thus the observer-frame rise time cannot be any longer
than 26~days.
% ROTSE rest-frame day -16   = Conley day -11,   detected & matched
% ROTSE rest-frame day -18.7 = Conley day -14.2, detected & matched
% P/Q   rest-frame day -26   =                   not detected
A parabolic \citep{riess99,akn00} fit to the first four points on the ROTSE
light curve \citep{cbet2007if} suggests an explosion date of 2007~August~10.7
(MJD~54322.6); this gives a rest-frame rise time of $24.2 \pm 0.4$~days,
which we adopt for our analysis.

Using these numbers we derive a \nickel\ mass of
$M_\mathrm{\nickel} = (1.72 \pm 0.09) \times (\alpha/1.2)~\Msol$,
somewhat larger than SN~2003fg and SN~2009dc, and already in excess of
the Chandrasekhar mass for $\alpha = 1.2$.
This estimate assumes \emph{no} extinction due to dust in the host galaxy.
The Chandrasekhar-mass DET1 model of \cite{kmh93}, in contrast, makes only
0.92 \Msol\ of \nickel, approximately the theoretical upper limit for
Chandrasekhar-mass detonations.
We will return to the \nickel\ mass estimate below in the context of more
detailed models of the SN~2007if progenitor.

% ############################################################################

\section{Constraints on the SN~2007if total mass and density structure}

There have been several attempts to explain how a
Chandrasekhar-mass white dwarf in a single-degenerate scenario could
give rise to an unusually luminous explosion, usually relying on departures
from spherical symmetry.  \cite{hsr07} simulated aspherical
three-dimensional SN~Ia explosions, which at their brightest achieved
$M_\mathrm{bol} = -19.9$ at maximum, but at the cost of some fine-tuning
and requiring an almost complete conversion of the star to \nickel.
They also failed to reproduce the low velocities found in SN~2003fg-like SNe.
\cite{kasen04} considered initially spherical explosions which lost spherical
symmetry through the interaction of the SN~ejecta with a non-degenerate
companion star, forming a conical hole through which deeper, hotter ejecta
became visible along certain lines of sight; such a model produces a SN with a
1991T-like spectrum, bluer colors and only a modest ($\sim 0.3$ mag)
luminosity enhancement relative to normal SNe~Ia.
Neither of these models are consistent with our observations of SN~2007if.

\cite{tanaka09} present polarimetry observations for SN~2009dc which show
no significant departures from spherical symmetry.  This would seem to imply
that, whether or not the progenitor star is aspherical, a highly aspherical
explosion is not necessary to attain a very high luminosity.  However, it is
reasonable to ask whether observations can provide further constraints on
the mass of the progenitors in such systems apart from measuring the SN's
luminosity and velocity near peak, within the context of an explosion with
no large-angular-scale anisotropies.  One way of approaching this question
is via the late-time bolometric light curve \citep{jeffery99,stritz06},
which requires an accurate estimate of the kinetic energy of the explosion
and the corresponding velocity scale for the ejecta density profile.

Our measurements (see Figure~\ref{fig:vsi-comp}) show an unusually low
\ion{Si}{2} $\lambda 6355$ velocity in the ejecta, as has been seen in
other super-Chandrasekhar SN~Ia candidates.  We also see a fairly flat
velocity evolution, consistent with a plateau in the \ion{Si}{2} velocity
\citep[$\delta v \leq 500~\mathrm{km~s^{-1}}$;][]{qhw07} starting
as early as day $-9$ and extending at least as late as day $+7$.
These low velocities have previously been attributed to large binding energies
for these events \citep{howell06}.  In the following section we explore
an alternative --- that the low velocity might result from deceleration of the
outer layers of ejecta by a massive envelope surrounding the SN progenitor,
producing an overdense shell in the ejecta.  We present a simple parametrized
treatment of the shell structure which allows us to constrain the true
underlying kinetic energy of the explosion, the mass of the shell and the
envelope, and the total ejected mass.

% ============================================================================

\subsection{Shell structure in SN~2007if}

% Narrative:
% -- Present the spectroscopic observations as evidence for a shell
% -- Present the Sobolev mass estimate for the shell
% -- Argue that the DET2ENVN models of KMH93 show an undisturbed exponential
%    density distribution underneath the shell
% -- Use this to relate the shell and progenitor masses to each other

\cite{qhw07} studied the influence of shell structure in SN~Ia ejecta for
a normal SN~Ia event, SN~2005hj.  They pointed out that an overdense spherical
shell in the ejecta expanding at a given velocity slows the recession of the
photosphere, producing a plateau at that velocity in the time evolution of
the widths of absorption features, e.g., \ion{Si}{2}.  The shell would also
be characterized by a duration and a velocity-space width from the spectra.
Photometrically, SNe~Ia with an overdense shell in the outer layers should
appear redder and brighter than SNe~Ia with no shell.
Existing models of shell-structure SNe~Ia have been calculated only for
Chandrasekhar-mass explosions, but one should expect
analogous physical effects in super-Chandrasekhar-mass explosions.
Such models provide an alternate interpretation for the low \ion{Si}{2}
velocities seen in SN~2003fg and SN~2007if, although a large progenitor binding
energy may also contribute.

Shell structure occurs naturally in DD merger scenarios, in which the
tidal disruption of one or both white dwarfs creates a common envelope of
carbon and oxygen around the central merger product.
Collision of the ejecta with the envelope then accelerates the envelope
to higher velocities and creates a strong reverse shock which creates a
shell in the outermost SN ejecta.
\cite{hk96} simulate several instances of common envelopes
in their DET2ENVN ``tamped detonation'' (TD) models, which place two merging
white dwarfs of total mass 1.2 solar masses in an envelope of mass
$0.1 \times N~\Msol$.
DET2ENV4 and DET2ENV6 produce $B-V \sim 0.2$ near maximum light,
significantly redder than SN~2006gz but consistent with 2007if.
The DET2ENV6 model, with a total system mass of 1.8 \Msol, also has
$\Delta m_{15}(V) = 0.63_{-0.09}^{+0.15}$, the slowest of any model in that
work, compared with $\Delta m_{15}(V) = 0.50 \pm 0.07$ for SN~2007if.
(The number for SN2007if is interpolated from the observations using
 the SALT2 fit; the dominant source of error for both the models and the
 observations comes from the uncertain phase of maximum light).

The single-degenerate pulsating delayed detonation (PDD) models of \cite{hk96}
also have shell structure in the ejecta.  Here an initial deflagration phase
pre-expands the white dwarf; the outermost layers begin to recollapse and
are then entrained into a shell by the underlying ejecta from a subsequent
detonation.  These models do exhibit an intrinsically red $B-V$ color
at maximum, though with a wide spread (0.05--0.60).
However, they have unusually short rise times (13--15~days)
and slower declines than observed for SN~2007if.  Most of the \cite{hk96}
models have rise times which are too short --- around 15~days in $V$, compared
to the observed typical value of 19~days \citep{conley06}.  This aspect of the
models could no doubt be improved for a more accurate match to observations.
Nevertheless, it is striking that the shell-structure explosion models reach
the extremes of the rise time distribution with fairly small dispersion,
with most PDD models at 13~days
and the TD models as large as 22~days.  The one PDD model that achieves
a sufficiently long rise time and slow decline, PDD535, is much fainter
even than normal SNe~Ia ($M_V = -17.77$) and has an extremely red color
of $B-V = 0.60$.  This model produces about 20\% of the kinetic energy
of the DET2ENVN models.

A separate question arises regarding the widespread use of the differentially
rotating models of \cite{yl05} in treatments of super-Chandrasekhar-mass
SNe~Ia \citep{howell06,jbb06,maeda09b}, including this paper
(see below).  \cite{piro08} found that inclusion of the baroclinic instability
and of magnetohydrodynamic processes in white dwarf interiors inhibits the
Kelvin-Helmholtz instabilities expected in differentially rotating white
dwarfs, limiting such white dwarfs to rigid rotation and apparently making
single-degenerate super-Chandrasekhar-mass white dwarfs untenable.
Even discounting such effects, however, supermassive white dwarfs accreting
from a non-degenerate companion are not expected to evolve to masses above
1.7 \Msol\ \citep{chenli09}; the progenitor of SN~2007if was almost certainly
more massive than this.

The interpretation of a velocity plateau as the sign of an interaction with
an envelope containing a few tenths of a solar mass of material also poses 
problems for models involving the explosion of single-degenerate supermassive
white dwarfs.  To the extent that such an envelope is necessary to explain
the kinematics of 2003fg-like SNe~Ia, it must be composed of carbon and
oxygen; a hydrogen/helium envelope of comparable mass would most likely
produce strong emission lines such as those seen in the ``SNe IIa''
2002ic \citep{hamuy03,mwv04} and 2005gj \citep{snf2005gj,sdss2005gj}.

The above considerations disfavor existing single-degenerate explosion models,
although it is conceivable that \emph{some} PDD model could be tuned to fit
our observations.  The TD models provide a better fit to the observations.
For further analysis below, we therefore discount models of pulsating delayed
detonations for the time being, and concentrate instead on tamped detonations
representing double-degenerate mergers, using DET2ENVN as our main point of
reference.

% ============================================================================

\subsection{Near-IR light curves and mixing of the ejecta}
\label{subsec:kasen-ir}

Our estimate of the total ejected mass will follow
\cite{jeffery99} and \cite{stritz06}, relying on the estimated transparency
of the ejecta to gamma rays from \cobalt\ decay.  The model requires an
assumption about the distribution of \nickel\ in the ejecta.  Although this
incurs some uncertainty, we argue that the ejecta of SN~2007if
are fairly well mixed based on the near-infrared (NIR) light curves.

\cite{kasen06} presents radiative transfer Monte Carlo calculations which
illustrate the relevance of mixing in SN~Ia ejecta to the appearance of an
inflection point or second maximum in their NIR light curves.
The recombination of iron provides an efficient means of
redistributing light from UV and blue bands to the NIR, and recombination
happens at a more or less sharp front in SN~Ia ejecta as they expand and cool.
Thus in SNe with compact cores of iron-peak elements, a wave of recombination
of \ion{Fe}{3} to \ion{Fe}{2} powers the first NIR light curve peak, while
the recombination of \ion{Fe}{2} to \ion{Fe}{1} powers the well-known
second peak.  Supernovae in which the ejecta are well-mixed have
less pronounced second maxima in the near infrared than those in which the
ejecta are stratified.  Supernovae which produce less \nickel\ also show less
pronounced second maxima, and these maxima occur earlier in phase and are
more closely spaced with respect to the first maximum, since the temperature
of the ejecta is lower and recombination of iron-peak elements occurs earlier.

The $I$-band light curve of 2007if shows only a very slight inflection which
is difficult to measure.  The $J$-band light curve shows stronger, but still
quantitatively weak, evidence for a second maximum.
In the context of this model, the observed $I$- and $J$-band light curves of
SN~2007if suggest either a low \nickel\ mass, ejecta which are highly mixed,
or both.  Assuming stratified ejecta, the shape, timing and duration of the
$J$-band second peak is broadly consistent with a large amount of \nickel\
(0.6 \Msol\ or more).  Given the luminosity of the event, a plausible
interpretation is that the ejecta contain a large amount of \nickel, much of
which lies near the surface --- either because it has been mixed to higher
velocities in the explosion, or because the ejecta are composed mostly of
\nickel\ and little else.  This interpretation is also supported by
the faster decline in $B$-band relative to SN~2003fg, suggesting that the
$B$-band flux is indeed being efficiently reprocessed to NIR wavelengths
\citep{kw07}; a smaller amount of \nickel\ might manifest as a sharper
decline in all bands
(i.e. the bolometric light curve) rather than $B$-band only.
It is also consistent with the appearance of \ion{Fe}{3} lines near maximum,
as mentioned in Section \ref{sec:spectra_obs}.

% chisq = 19.8/(10-2) = 2.5, excludes straight-line fit at 99% CL
% chisq = 15.6/(10-2) = 2.2, excludes parabolic fit at 95% CL
% nicely consistent with 4th-order polynomial; so 2nd max probably real

\cite{kmh93} point out that in SNe with shell structure, the shock interface
between the shell and the high-velocity ejecta is Rayleigh-Taylor unstable.
This could provide a mechanism
for mixing of iron-peak elements into the shell in SN~2007if.
If so, we might expect the associated features --- a weak NIR second maximum
and a relatively sharp B-band decline rate --- to be a generic feature even
of shell-structure SNe with less massive progenitors.  We might therefore
also expect the usual width-luminosity relationship to break down for
shell-structure SNe~Ia, producing SNe which are systematically
brighter for their ($B$-band) decline rate than SNe~Ia with near-exponential
density structures \citep{kw07}.  Whether the bolometric light curves still
followed a width-luminosity relationship would depend in part on the
efficiency of the optical-NIR reprocessing.

% RS:  Dan goes on to talk about the Ca II triplet feature, and points out
% that if they are treated as purely absorbing, they can raise the I-band
% magnitude by up to 0.5 mag and delay the first peak by a week, as they
% redistribute flux from B-band.  He doesn't think this is what happens in
% nature, though.  Otherwise it would be empirically be hard to explain the
% good fits pure scattering models provide to real SNe Ia, and the observed
% low scatter of SN light curves in the infrared.

% ============================================================================

\subsection{Mass estimates and photospheric evolution
            for SN~Ia models with exponential density profiles}
\label{sec:mass-eject}

In the optically thin limit, and when the gamma rays come from a single
radioactive species, Equation (\ref{eqn:opacity}) is modified by multiplying
the $Q_\mathrm{Co,\gamma}$ term by a factor $1 - \exp(-\tau_\gamma)$,
where $\tau_\gamma$ is the optical depth to scattering of gamma rays
from the decay of \cobalt.  Such approximations are useful in describing
the state of the ejecta by day $+60$ after explosion
\citep{jeffery99,stritz06}, by which time less than 0.5\% of the initial
\nickel\ should remain.

The optical depth $\tau_\gamma$ should scale as $t^{-2}$ for SN~ejecta in
homologous expansion.  The evolution of the ejecta density in a SN~Ia
can be well described by
\begin{equation}
\rho(v,t) = \rho_\mathrm{c,0} (t_0/t)^3 \exp (-v/v_e),
\end{equation}
where $\rho_\mathrm{c,0}$ is the central density at fiducial time $t_0$
and $v_e$ is the scaling velocity.
By taking $t_0$ to be the time at which $\tau_\gamma = 1$,
i.e. the transition to the optically thin regime \citep{jeffery99}
and integrating outwards from the center of the SN, we obtain
$\rho_\mathrm{c,0} = (\kappa_\gamma q v_e t_0)^{-1}$,
where $\kappa_\gamma$ is the opacity for gamma-ray absorption, and $q$ is a
form factor describing how the \nickel\ is distributed throughout the ejecta.
The total mass is
\[ M_\mathrm{WD} = 8\pi \rho_\mathrm{c,0} (v_e t_0)^3, \]
and the kinetic energy is
\[ E_K = 6 M_\mathrm{WD} v_e^2. \]
We can then use these relations to estimate the total mass as
\begin{equation}
M_\mathrm{WD} = \frac{8\pi}{\kappa_\gamma q} (v_e t_0)^2.
\end{equation}
In their analysis of a sample of normal SN~Ia bolometric light curves,
\cite{stritz06} adopt $q = 0.33 \pm 0.09$
(appropriate for ejecta in which \nickel\ is mixed evenly throughout),
$\kappa_\gamma = 0.025 \pm 0.03~\mathrm{g~cm}^{-2}$,
and $v_e = 3000$~km~s$^{-1}$.
They emphasize that the procedure tends to produce low ejected masses
($< 1.0~\Msol$) for many SNe~Ia, and comment on the accuracy of the input
parameters.  The values of $\kappa_\gamma$ and $q$ are not expected to vary
much, and the largest single uncertainty is expected to arise from the value
of $v_e$ used, which may have uncertainties on the order of 20\% and is
squared in the above expression.

For ejecta following an exponential density distribution, the photospheric
velocity $v_\mathrm{phot}$ can also be calculated given $v_e$ by solving
$\tau_\mathrm{opt}(v_\mathrm{phot}) = 2/3$ ($\sim 50\%$ escape probability).
This leads to
\begin{equation}
v_\mathrm{phot} = v_e \, \ln
   \left( \frac{3 \kappa_\mathrm{opt} M_\mathrm{WD}}{16\pi v_e^2 t^2} \right)
\end{equation}
where $\kappa_\mathrm{opt}$ is now the opacity for optical photons.  For a
near-exponential density profile, we would expect \ion{Si}{2} $\lambda 6355$
to track the photosphere (and hence $v_e$) for observations near maximum
light; this is the assumption made by \cite{maeda09b} in their parameter
study of super-Chandrasekhar-mass SNe~Ia.

% ============================================================================

\subsection{Incorporating the effects of a shell}

The ejected mass estimate in Section \ref{sec:mass-eject} hinges on the
assumption that the ejecta follow an exponential density distribution with
velocity.  This is no longer true for SNe with shell structure.  However,
the DET2ENVN models have an essentially undisturbed exponential
density structure in the centrally concentrated ejecta underneath the shell.
Although the presence of a shell modifies the velocity structure and colors
of the SN considerably, it should have negligible effect on the bolometric
light curve at sufficiently late times, since the opacity $k_\gamma$ is an
order of magnitude smaller than $k_\mathrm{opt}$ in most realistic scenarios.
Due to geometrical dilution, the shell becomes transparent to gamma rays by
maximum light, after which the exponential models should apply.
The shell mass should consist simply of the ejected mass above some
velocity $v_\mathrm{sh}$, which looks like
\begin{equation}
\frac{M(v>v_\mathrm{sh},t)}{M_\mathrm{WD}}
   = Q \left(3,\frac{v_\mathrm{sh}}{v_e}\right),
\end{equation}
where $Q(a,x) = \gamma (a,x) / \Gamma(a)$ is the incomplete gamma function.
Similar gamma-function integrals can be calculated to give
the \emph{fractions} of the optical depth, momentum, and kinetic energy,
for which the values of $a$ are 1, 4, and 5, respectively.  For a typical
model with $v_e = 2750$~km~s$^{-1}$, the material above 9000~km~s$^{-1}$
carries about 35\% of the ejected mass and 75\% of the kinetic energy,
yet contributes less than 5\% to the gamma-ray optical depth as measured from
the center of the explosion to infinity.

We therefore expect that the bolometric light curve will yield a reasonable
mass estimate for SN~2007if, provided we choose $v_e$ appropriately.
However, in the presence of a shell, the \ion{Si}{2} velocity is \emph{not}
representative of the kinetic energy velocity $v_e$ during the velocity
plateau, but instead represents the shell velocity.
Our observations show no significant change in the \ion{Si}{2} velocity
over a 16-day period ($-9$~days to $+7$~days), inconsistent with
the photospheric evolution of a near-exponential density profile even
for SNe~Ia with very low kinetic energy.  Although \ion{Si}{2} $\lambda 6355$
is sometimes used at late times ($+40$~days) to estimate $v_e$, the line
is too heavily blended with other features in the SN~2007if post-plateau
spectra to measure its velocity.  We therefore cannot observe $v_e$ directly
in our work here, and instead marginalize over it as a nuisance parameter.

For double-degenerate progenitor models with envelopes, we can derive
the shell and envelope masses given $v_e$ via
conservation of momentum and optical transparency arguments.  Shortly after
explosion, all SN ejecta above $v = v_\mathrm{sh}$ are coasting freely and
the envelope with mass $M_\mathrm{env}$ is stationary.  At late times, the
envelope has been accelerated to higher velocities, and the outermost SN
ejecta are moving together in a shell of negligible thickness with velocity 
$v_\mathrm{sh}$ (which will match our measured \ion{Si}{2} velocities).
In the DET2ENVN models, the specific
momentum in the C/O envelope is some 50\% higher than in the shell.
Conservation of momentum then gives
\begin{eqnarray}
p_\mathrm{initial}
  & = & 3 M_\mathrm{WD} v_e Q \left(4,\frac{v_\mathrm{sh}}{v_e}\right)
   = p_\mathrm{final} \nonumber \\
  & = & \left[M_\mathrm{WD} Q \left(3,\frac{v_\mathrm{sh}}{v_e}\right)
    + 1.5 M_\mathrm{env}\right] v_\mathrm{sh}.
\end{eqnarray}
Some algebra then gives
\begin{equation}
M_\mathrm{env} = \frac{2}{3} \left[ \frac{3 v_e}{v_\mathrm{sh}}
     Q\left(4,\frac{v_\mathrm{sh}}{v_e}\right)
   - Q\left(3,\frac{v_\mathrm{sh}}{v_e}\right) \right]
   M_\mathrm{WD}.
\label{eqn:envmass}
\end{equation}
The sum $M_\mathrm{tot} = M_\mathrm{env} + M_\mathrm{WD}$ should then equal
the mass of the two white dwarfs in the original close binary system.
The combined envelope-plus-ejecta shell should remain opaque for a time
\begin{equation}
t_\mathrm{sh} = \frac{1}{v_\mathrm{sh}}
   \sqrt{\frac{\kappa_\mathrm{opt} M_\mathrm{sh}}{4\pi}}.
\label{eqn:tsh}
\end{equation}
We can calculate $t_\mathrm{sh}$, assuming fractions $f_\mathrm{IPE}$
and $f_\mathrm{IME}$ of iron-peak (Ni, Co, Fe) and intermediate-mass elements
(including carbon and oxygen) respectively, by using the approximate
line-opacity prescription of \cite{mazzali01},
\begin{equation}
\kappa_\mathrm{opt} = 0.5 \left[
   0.25 f_\mathrm{IPE} + 0.025 f_\mathrm{IME}
   \right]~\mathrm{cm^2~g^{-1}}
\label{eqn:kopt}
\end{equation}
\citep[see also][]{maeda09b}.  Our use of this relation assumes the material
in the shell has been mixed.  The conserved momentum and the observed
transparency of the shell provide indirect constraints on the underlying,
unobserved $v_e$.

% ----------------------------------------------------------------------------

\subsection{Confidence regions on the progenitor mass}
\label{sec:confint-fiducial}

\begin{deluxetable*}{lccccccccr}
\tabletypesize{\footnotesize}
\tablecaption{Variation in constraints with model-dependent parameters}
\tablehead{
   \colhead{$\alpha$} &
   \colhead{$E(B-V)_\mathrm{host}$} &
   \colhead{$R_{V,\mathrm{host}}$} &
   \colhead{$\rho_c$~($10^9$~g~cm$^{-3}$)} &
   \colhead{$M_\mathrm{tot}~(\Msol)$} &
   \colhead{$M_\mathrm{WD}~(\Msol)$} &
   \colhead{$M_\mathrm{Ni}~(\Msol)$} &
   \colhead{$f_\mathrm{env}$} &
   \colhead{$f_\mathrm{sh}$} &
   \colhead{Prob.\tablenotemark{a}}
}
\startdata
% RS note to self -- check probabilities of non-shell solutions for high alpha
0.80 & 0.00 & --- & 3.00
     &  --- & --- & --- % & $2.23 \pm 0.03$ & $1.95 \pm 0.02$ & $1.90 \pm 0.02$
     &  --- & ---       % & $0.14 \pm 0.01$ & $0.42 \pm 0.01$
     & $< 10^{-5}$ \\ % (0.00464)
0.90 & 0.00 & --- & 3.00
     & $2.26 \pm 0.06$ & $2.00 \pm 0.06$ & $1.84 \pm 0.06$
     & $0.14 \pm 0.01$ & $0.39 \pm 0.02$ & 0.004 \\ % (1.512)
1.00 & 0.00 & --- & 3.00
     & $2.28 \pm 0.09$ & $2.03 \pm 0.08$ & $1.78 \pm 0.07$
     & $0.14 \pm 0.02$ & $0.38 \pm 0.02$ & 0.078 \\ % (34.88)
1.10 & 0.00 & --- & 3.00
     & $2.32 \pm 0.12$ & $2.08 \pm 0.11$ & $1.71 \pm 0.09$
     & $0.15 \pm 0.03$ & $0.37 \pm 0.02$ & 0.378 \\ % (168.8)
1.20 & 0.00 & --- & 3.00
     & $2.36 \pm 0.15$ & $2.13 \pm 0.13$ & $1.63 \pm 0.09$
     & $0.15 \pm 0.03$ & $0.36 \pm 0.03$ & 0.798 \\ % (355.8)
1.30 & 0.00 & --- & 3.00
     & $2.41 \pm 0.17$ & $2.18 \pm 0.15$ & $1.55 \pm 0.09$
     & $0.16 \pm 0.04$ & $0.34 \pm 0.03$ & 1.000 \\ % (446.0)
1.40 & 0.00 & --- & 3.00
     & $2.45 \pm 0.18$ & $2.23 \pm 0.15$ & $1.48 \pm 0.08$
     & $0.16 \pm 0.04$ & $0.32 \pm 0.03$ & 0.836 \\ % (373.0)
1.50 & 0.00 & --- & 3.00
     & $2.49 \pm 0.18$ & $2.28 \pm 0.16$ & $1.41 \pm 0.08$
     & $0.17 \pm 0.05$ & $0.31 \pm 0.03$ & 0.488 \\ % (217.8)
1.60 & 0.00 & --- & 3.00
     & $2.53 \pm 0.17$ & $2.33 \pm 0.15$ & $1.34 \pm 0.07$
     & $0.18 \pm 0.05$ & $0.30 \pm 0.03$ & 0.208 \\[1.0mm] % (92.67)
\hline \\ % ------------------------------------------------------------------
1.30 & 0.09 & 2.1 & 3.00
     & $2.42 \pm 0.11$ & $2.17 \pm 0.09$ & $1.87 \pm 0.09$
     & $0.15 \pm 0.02$ & $0.36 \pm 0.03$ & 0.546 \\ % (9.86/18.07)
1.30 & 0.09 & 3.1 & 3.00
     & $2.45 \pm 0.09$ & $2.20 \pm 0.08$ & $1.98 \pm 0.10$
     & $0.15 \pm 0.02$ & $0.36 \pm 0.03$ & 0.231 \\ % (4.17/18.07)
1.30 & 0.18 & 2.1 & 3.00
     & $2.50 \pm 0.06$ & $2.25 \pm 0.06$ & $2.03 \pm 0.10$
     & $0.14 \pm 0.02$ & $0.36 \pm 0.03$ & 0.019 \\ % (0.350/18.07)
1.30 & 0.18 & 3.1 & 3.00
     &  --- & --- & --- % & $2.67 \pm 0.04$ & $2.38 \pm 0.03$ & $2.32 \pm 0.03$
     &  --- & ---       % & $0.12 \pm 0.01$ & $0.37 \pm 0.01$
     & $< 10^{-5}$ \\[1.0mm] % (0.002/1493.0)
\hline \\ % ------------------------------------------------------------------
1.30 & 0.00 & --- & 0.03
     & $2.72 \pm 0.07$ & $2.40 \pm 0.06$ & $1.61 \pm 0.08$
     & $0.39 \pm 0.03$ & $0.40 \pm 0.03$ & 0.213 \\ % (95.11/446.0)
1.30 & 0.00 & --- & 0.10
     & $2.66 \pm 0.11$ & $2.36 \pm 0.09$ & $1.61 \pm 0.09$
     & $0.31 \pm 0.04$ & $0.39 \pm 0.03$ & 0.631 \\ % (281.6/446.0)
1.30 & 0.00 & --- & 0.30
     & $2.62 \pm 0.14$ & $2.33 \pm 0.11$ & $1.60 \pm 0.09$
     & $0.25 \pm 0.04$ & $0.38 \pm 0.03$ & 1.044 \\ % (465.7/446.0)
1.30 & 0.00 & --- & 1.00
     & $2.55 \pm 0.16$ & $2.29 \pm 0.13$ & $1.58 \pm 0.09$
     & $0.20 \pm 0.04$ & $0.37 \pm 0.03$ & 1.332 % (594.4/446.0)
\enddata
\label{tbl:mlim-alpha-vary}
\tablecomments{Constraints on properties of models % with $f_\mathrm{CO} = 0$
as a function of the (model-dependent) parameters
$\alpha = L_\mathrm{bol}/L_\mathrm{rad}$, $E(B-V)_\mathrm{host}$,
$R_{V,\mathrm{host}}$, and $\rho_c$.
Columns, left to right:  $\alpha$; assumed host extinction $E(B-V)$ and $R_V$;
assumed central density $\rho_c$ of the white dwarf before explosion;
mass of the white dwarf merger product
$M_\mathrm{WD}$; \nickel\ mass synthesized in the explosion;
ratio of envelope mass to white dwarf mass; ratio of shell mass to
white dwarf mass; total probability, marginalized over all free parameters.}
\tablenotetext{a}{Normalized to $\alpha = 1.3$, no host extinction,
$\rho_c = 3 \times 10^9$~g~cm$^{-3}$ (fiducial analysis).}
\end{deluxetable*}

In our analysis, following \cite{jeffery99},
we adopt $\kappa_\gamma = 0.029$~cm$^2$~g$^{-1}$ and $q = 0.33$
(since an even distribution of \nickel\
is supported by our observed NIR light curves).  The appropriate value for
$v_e$ will depend on the kinetic energy, which in turn will depend on the
mass and on the composition of the ejecta.  To get a self-consistent mass
estimate, we generate a suite of semi-analytic progenitor models similar
to those presented in \cite{jbb06}, but including an envelope with
characteristics similar to those found in DET2ENVN.
We parametrize our models by
the white dwarf mass $M_\mathrm{WD}$ and the fractions $f_i$ of \nickel,
other iron-peak elements (``Fe''), intermediate-mass elements
(``Si'') and unburned carbon and oxygen (``C/O''),
the radioactive energy deposition parameter $\alpha$,
and the ratio $f_\mathrm{env}$ of the C/O envelope mass $M_\mathrm{env}$
to the ejected mass $M_\mathrm{WD}$ which contributes to the original binding
energy.  The binding energy was calculated from the
differentially rotating white dwarf models of \cite{yl05}, and the kinetic
energy was calculated as the difference between the binding energy and the
nuclear energy released.

After calculating $v_e$ for each model, we could then make predictions for
the observed bolometric light curve to be compared directly with observations.
In matching the bolometric luminosity at peak (Arnett's rule) jointly with the
other observations, we sample values of $\alpha$ between 0.8 and 1.6 to cover
a reasonable range of potential variation in different explosion scenarios.
However, for our primary analysis we adopt a fiducial value of $\alpha = 1.3$,
characteristic of the tamped-detonation models DET2ENVN which lie within
$\pm 0.05$ of this value \citep{hk96}.
We calculate $v_\mathrm{sh}$ given $v_e$ and $M_\mathrm{env}$ by solving
Equation (\ref{eqn:envmass}) numerically, and we calculate $t_\mathrm{sh}$
from equations \ref{eqn:tsh} and \ref{eqn:kopt}
(using $f_\mathrm{IPE} = f_\mathrm{Fe} + f_\mathrm{Ni}$
   and $f_\mathrm{IME} = f_\mathrm{Si} + f_\mathrm{CO}$).
In constraining the models we require that $v_\mathrm{sh}$ match
the observations ($9000 \pm 500$~km~s$^{-1}$) and that $t_\mathrm{sh}$
be longer than the observed lifetime of the plateau
($+7$~days after $B$-band maximum, or $+32$~days after explosion).
Although the spectral characteristics begin to change at day $+10$ after
$B$-band maximum, including a weakening of \ion{Si}{2} $\lambda 6355$
which signals the end of the plateau phase, we do not enforce a hard
upper limit on the length of the plateau.

The value of $M_\mathrm{tot}$ was sampled uniformly between 1.4 and 2.8 \Msol,
with the highest allowed mass corresponding to the merger of two
Chandrasekhar-mass white dwarfs.  We chose $2~\mch$ as the most conservative
upper bound on the system mass, since observations, or the theory of
common-envelope evolution thought to produce double-degenerate systems,
do not yet provide secure constraints
\citep[see for example][]{dobbie06,dobbie09}.
The values $f_\mathrm{Fe}$, $f_\mathrm{Ni}$, $f_\mathrm{Si}$, and
$f_\mathrm{CO}$ (representing the composition of the bound progenitor mass)
were sampled uniformly between 0 and 1 such that their sum was unity.
The value of $f_\mathrm{env} = M_\mathrm{env}/M_\mathrm{WD}$ was varied
independently between 0 and 0.5.
The central density of the merged progenitor was fixed at
$3 \times 10^9~\mathrm{g~cm^{-3}}$, higher than the DET2ENVN models but typical
of central densities of white
dwarfs assumed elsewhere \citep{howell06,maeda09b}.
For models with $f_\mathrm{env} > 0$, i.e., models with shells,
we assume that all of the IME are mixed into the shell, to agree with
our observations of very weak or absent \ion{Si}{2} in post-plateau spectra.

To generate confidence regions, we evaluated a chi-square ($\chi^2$)
for each progenitor model when compared to all available observations.
Each observation --- the maximum bolometric luminosity, bolometric luminosity
measurements more than 60~days after maximum, and velocity measurements
during the plateau phase --- contributed equally to $\chi^2$ according
to its error.  Each model was then weighted by the probability of observing
the calculated value of $\chi^2$, and binned in a histogram.
The confidence regions drawn contain the stated fractions of the total
probability of all models calculated.  There may be additional constraints
on the actual distribution of these parameters realized in nature,
which we do not include in the
sampling or weighting of different possible combinations of parameters.
For example, SN~Ia explosion physics should constrain the relative
abundances of various elements in the ejecta; our constraints rely only
on the kinematics of the explosion given the nuclear energy released.
Due to this and other potential limitations,
we therefore do not claim to pick out a single set of values which is most
likely, nor to reproduce the detailed shapes of the distributions of
underlying quantities.  However, this approximate method should allow us to
rule out untenable models.  The size of the confidence regions, interpreted
loosely, should reflect the uncertainty on the measurements themselves.

\begin{figure*}
\center
\resizebox{\textwidth}{!}{\includegraphics{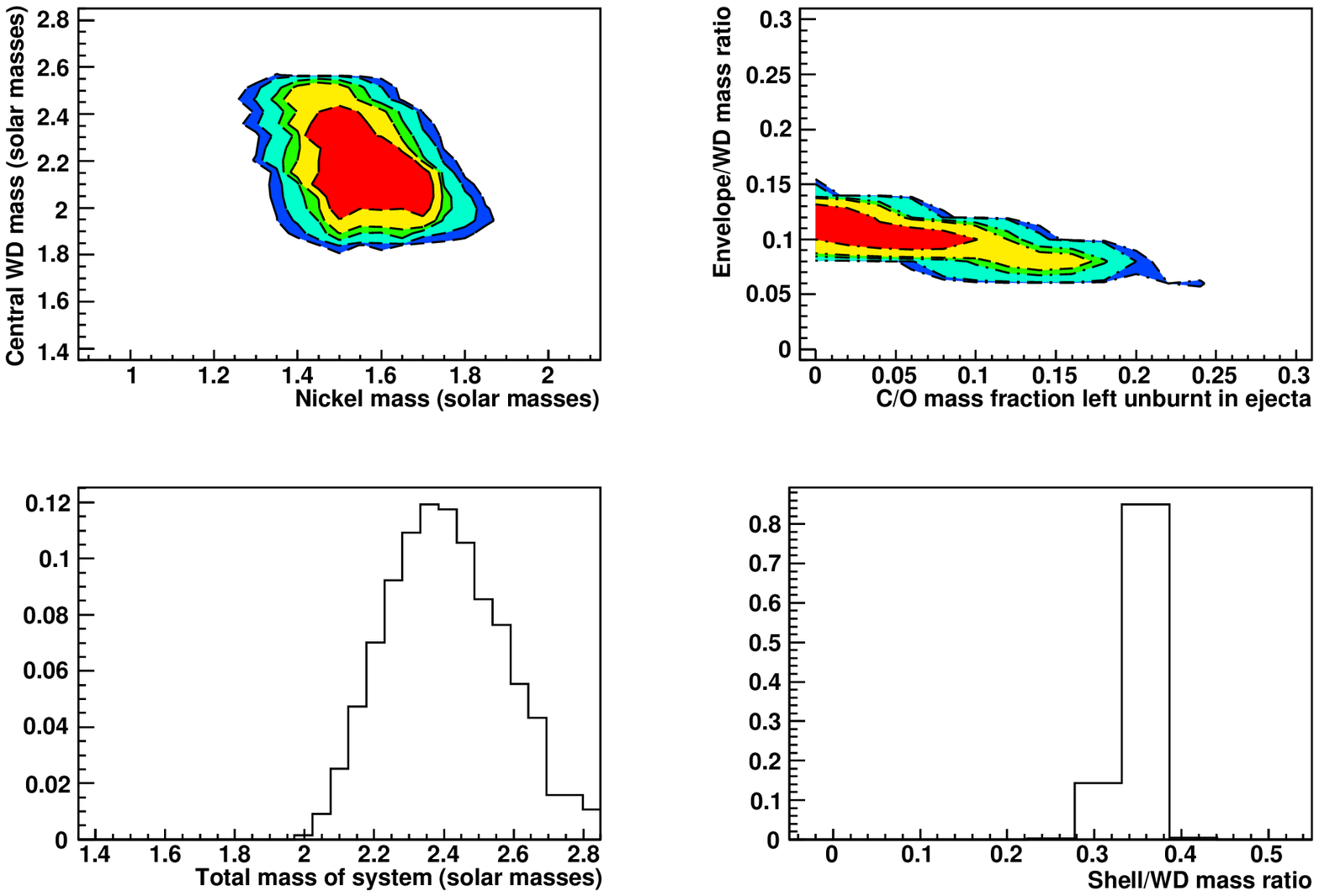}}
\caption{\small Constraints on progenitor models for SN~2007if,
assuming $\alpha = 1.3$ as for DET2ENVN
and allowing for a floating fraction of unburned carbon/oxygen
mixed into the shell of reverse-shock material.
Colored regions are 68\% (red), 90\%, 95\%, 99\%, and 99.7\% CL (blue).
Top left:  white dwarf mass $M_\mathrm{WD}$ vs. nickel mass $M_\nickel$
in solar masses.  Top right:  envelope mass fraction $f_\mathrm{env}$
vs. carbon-oxygen fraction $f_\mathrm{CO}$.
Bottom left:  probability distribution for the total mass
$M_\mathrm{tot} = M_\mathrm{WD} + M_\mathrm{env}$.  Bottom right:
probability distribution for the fraction of ejected mass
$f_\mathrm{sh} = M_\mathrm{sh}/M_\mathrm{WD}$ of material in the
dense shell of reverse-shocked ejecta.}
\label{fig:cont-env}
\end{figure*}

Figure~\ref{fig:cont-env} shows the first-pass results of this procedure
for the case $\alpha = 1.3$.
The constraints are surprisingly tight:  the data are consistent with a
model containing $2.18 \pm 0.15~\Msol$ of central ejected material, with
$1.55 \pm 0.09~\Msol$ of the mass being \nickel, wrapped in a C/O envelope
of mass at least $0.3~\Msol$.  The shell of reverse-shocked SN~ejecta
contains about a third of the central ejected mass, or some 0.7--0.9
\Msol\ of material.  The minimum chi-square value is $\chi^2_\nu = 25.1/27
= 0.93$, indicating that the model is able to satisfy all available
constraints.  Chandrasekhar-mass progenitors, and progenitors with no
envelopes, are ruled out at high significance.  We place a lower limit of
2.05 \Msol\ (99\% CL) on the total mass of the system, and 0.1 \Msol\
at similar confidence on the mass of the envelope.

The upper-right-hand plot in Figure~\ref{fig:cont-env} illustrate the
trade-off between unburnt, initially bound carbon/oxygen in the SN~ejecta
and more loosely bound carbon/oxygen in the surrounding envelope.
An envelope turns out to be much more effective than bound unburnt material
in producing the low velocities observed in the SN.  The slope of the contour
shows that about 4 times as much bound material is required to remain
unburnt (decreasing the released nuclear energy for a given binding energy)
as is needed to produce the same effect in an envelope
(decelerating initially fast-moving ejecta in a collision).
The inferred shell/envelope mass is therefore relatively insensitive to the
amount of material left unburned in the supernova explosion.

\begin{figure*}
\center
\resizebox{\textwidth}{!}{\includegraphics{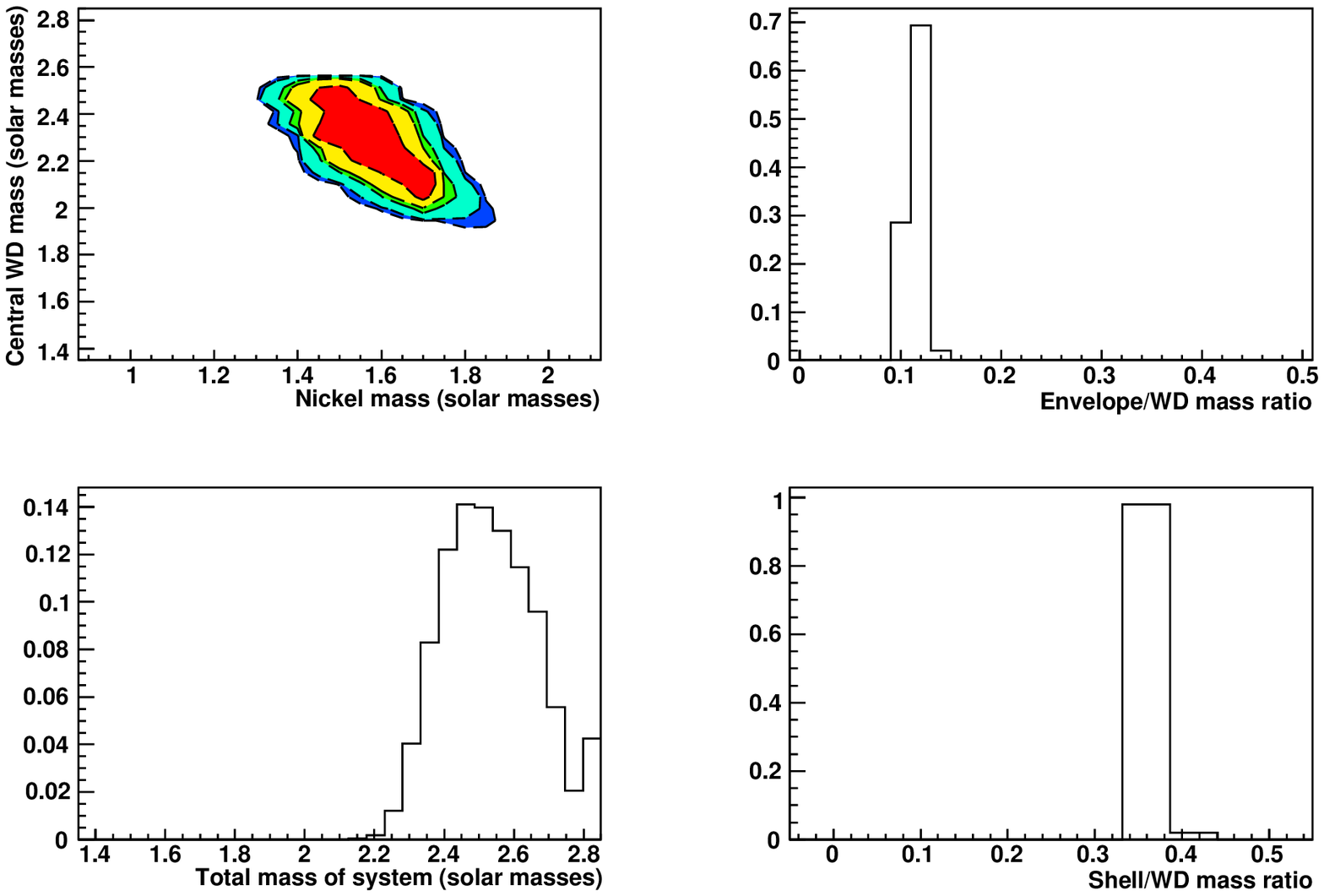}}
\caption{\small Constraints on progenitor models for SN~2007if,
assuming $\alpha = 1.3$ as for DET2ENVN
and assuming that all carbon/oxygen must reside
in the loosely bound envelope.
Colored regions are 68\% (red), 90\%, 95\%, 99\%, and 99.7\% CL (blue).
Top left:  white dwarf mass $M_\mathrm{WD}$ vs. nickel mass $M_\nickel$
in solar masses.  Top right:  probability distribution for the envelope
mass fraction $f_\mathrm{env}$.
Bottom left:  probability distribution for the total mass
$M_\mathrm{tot} = M_\mathrm{WD} + M_\mathrm{env}$.  Bottom right:
probability distribution for the fraction of ejected mass
$f_\mathrm{sh} = M_\mathrm{sh}/M_\mathrm{WD}$ of material in the
dense shell of reverse-shocked ejecta.}
\label{fig:cont-env-run2}
\end{figure*}

Given that \ion{C}{2} $\lambda 6580$ is the only visible carbon line in the
velocity-plateau spectra around $B$-band maximum, and is detected only
weakly, it seems unlikely that a great deal of unburned carbon/oxygen
is mixed into the shell.  Our observations can probably be adequately
reproduced by setting $f_\mathrm{CO} = 0$, so that all the unburned
material is required to lie in the envelope (which must then have a mass
of around 0.3 \Msol).  This is more in line with the hydrodynamic behavior
of the DET2ENVN models, and with our
understanding of the physics of normal SNe~Ia, in which almost all of the
material in the original white dwarf is burned in the explosion.
Figure~\ref{fig:cont-env-run2} shows how the mass constraints change when
we set $f_\mathrm{CO} = 0$; the low-\nickel\ models composed mainly of
nickel and carbon/oxygen are excluded here.
When the whole of $M_\mathrm{WD}$ must be burnt, much more additional mass
must be included to increase the binding energy sufficiently to reproduce
the appropriate value of $v_e$.
Under these assumptions, the 99\% CL lower limit on the system mass
shifts upward to 2.20 \Msol.

The results of varying $\alpha$ are shown in Table \ref{tbl:mlim-alpha-vary}.
The allowed region of parameter space shows a tight anticorrelation between
the central white dwarf mass and the \nickel\ mass.  This intriguing behavior
appears to result from the shape of the bolometric light curve:
models with lower \nickel\ mass favor higher values of $t_0$, i.e.,
the gamma-ray escape fraction at a given time is lower in these models.
It is interesting that the data seem to prefer $\alpha = 1.3$ even with no
outside input, providing further confirmation that this is a good choice
for interpretation of the data.

% ----------------------------------------------------------------------------

\subsection{The impact of uncertain reddening}
\label{sec:reddening}

The analysis in the previous section was carried out assuming
$E(B-V)_\mathrm{host} = 0$.
If the SN~is reddened by dust, we might expect the luminosity and the
\nickel\ mass to be higher, and to produce a higher total mass.  The other
observations limit the extent of this effect, since an increased mass
also leads to changes in the kinetic energy and density structure.
Given the hard upper mass limit of 2 \mch\ in our modeling,
applying an uncertain reddening correction has little impact on our
final mass constraints, but if certain reddening scenarios are shown to
be incompatible with our observations, we can rule them out,
even without recourse to the upper limit on \ion{Na}{1}~D absorption.

\begin{figure*}
\center
\resizebox{\textwidth}{!}{\includegraphics{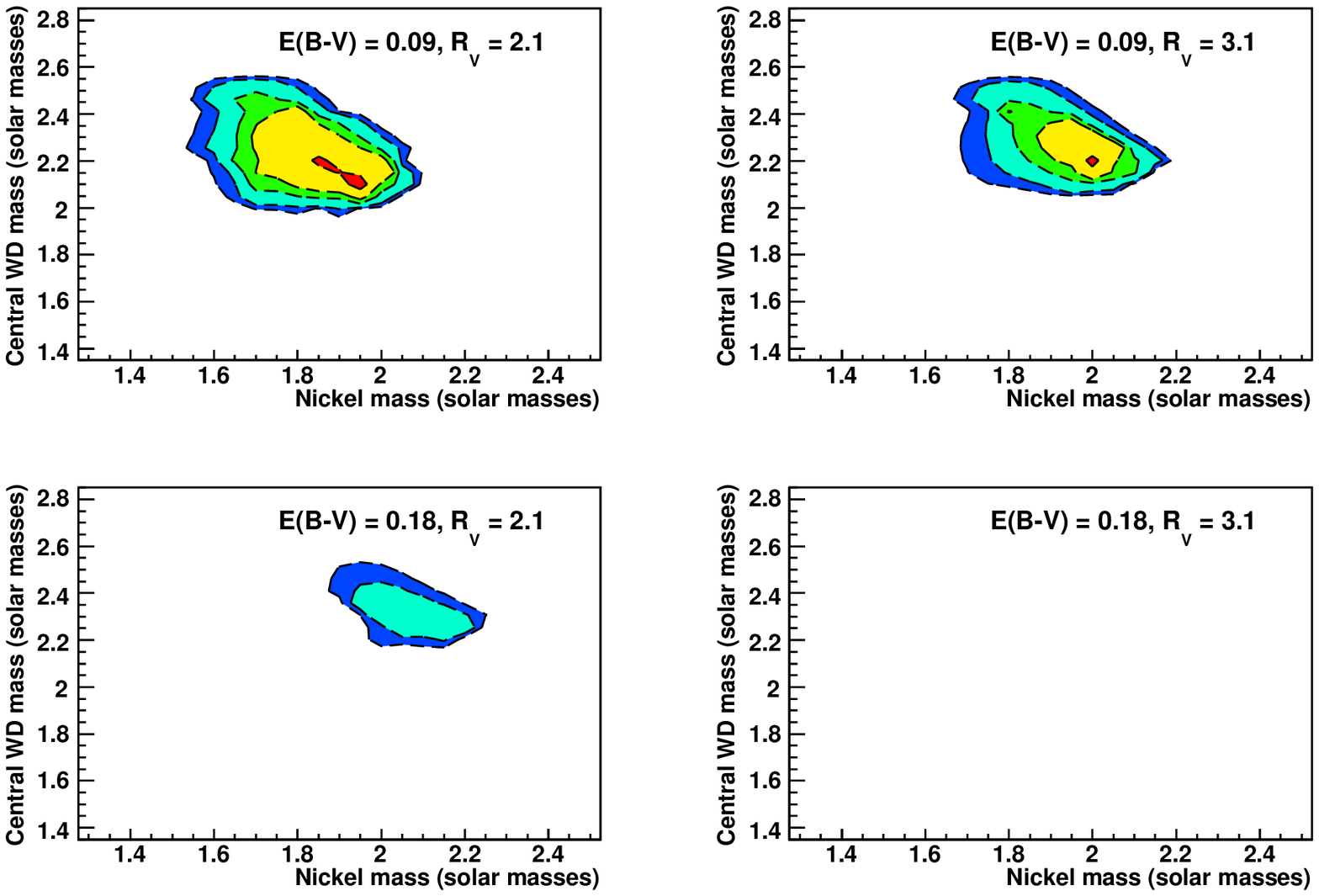}}
\caption{\small Constraints on progenitor model masses for SN~2007if
assuming $\alpha = 1.3$
(see e.g. upper left panel of Figure~\ref{fig:cont-env})
for different possible reddening scenarios.
Probability density is normalized to the zero-reddening case.
Colored regions are 68\% (red), 90\%, 95\%, 99\%, and 99.7\% CL (blue),
Top left:  $E(B-V) = 0.09$, $R_V = 2.1$.
Top right:  $E(B-V) = 0.09$, $R_V = 3.1$.
Bottom left:  $E(B-V) = 0.18$, $R_V = 2.1$.
Bottom right:  $E(B-V) = 0.18$, $R_V = 3.1$.}
\label{fig:cont-env-ebmv}
\end{figure*}

To constrain possible reddening scenarios, we repeated the analysis of
Section \ref{sec:confint-fiducial} for the case $\alpha = 1.3$, generating new
bolometric light curves which had been reddened according to various
prescriptions.  The results are shown in Figure~\ref{fig:cont-env-ebmv}.
These contours represent, in effect, different slices through the probability
density in parameter space.
We normalize the probability density displayed in each contour plot to the
total probability in the zero-reddening case.
This allows us to see how the relative total number of grid points
which satisfy the observations changes as the reddening is varied.
For example, the 68\% contour in each of the plots is set at a level which
encloses 68\% of the probability in the zero-reddening case.

We find that increasing degrees of reddening shift the allowed contours
to higher \nickel\ mass, as expected, although the total mass is not
strongly affected.
Our observations indicate $A_V < 0.38$ (99\% CL), and are therefore
incompatible with cases in which the Lira relation holds with
$E(B-V)_\mathrm{host} = 0.18$ and either $R_V = 2.1$ or $R_V = 3.1$.
High extinction tends to restrict solutions to a very small region of
parameter space, in which the ejecta are composed almost entirely
of \nickel.  Intermediate cases with $E(B-V) = 0.09$ are consistent with
the data and possibly also marginally consistent with the Lira relation
at the extremes of systematic deviation within the set for which it was
validated \citep{csp09}.

We conclude that, while some of the red color of SN~2007if may be put down
to dust, it is likely that the Lira relation overestimates the extinction,
although the slope of the color evolution of SN~2007if is similar to normal
SNe~Ia in the range of light curve phases where the Lira relation holds.
If true, this has impact for the other observed super-Chandrasehkar-mass
SN~Ia candidates in which the Lira relation is used to correct for reddening,
producing large uncertainties in the mass estimate from the SN~luminosity
alone for these SNe.

% ----------------------------------------------------------------------------

\subsection{The impact of uncertain central density}
\label{sec:density}

One can also vary the central density $\rho_c$ of the SN~progenitor at
ignition, another input which we held fixed in our calculations above.
The central density for carbon ignition in a Chandrasekhar-mass white dwarf
is expected to be around $3 \times 10^9~\mathrm{g~cm^{-3}}$ \citep{jbb06},
and is our fiducial value in the above analysis.
\cite{maeda09b} examine central densities
as high as $10^{10}~\mathrm{g~cm^{-3}}$, above which electron capture
is expected to make collapse to a neutron star more likely than
a SN~Ia \citep{nk91,yl05}.  The DET2ENVN models, which form
the primary point of reference for our analysis in the context of a tamped
detonation, actually have a much smaller central density, around
$4 \times 10^7~\mathrm{g~cm^{-3}}$.

There is therefore a dynamic range of at least two and a half orders of
magnitude within which $\rho_c$ could be expected to vary.
As with reddening, decreasing the central density can only increase the
total reconstructed mass (by decreasing the specific binding energy),
and so we expect it to have little impact on our final mass constraints.
However, by seeing how our observations constrain allowed masses in
different central density scenarios, we may be able to place a lower limit
on the allowed density.

\begin{figure*}
\center
\resizebox{\textwidth}{!}{\includegraphics{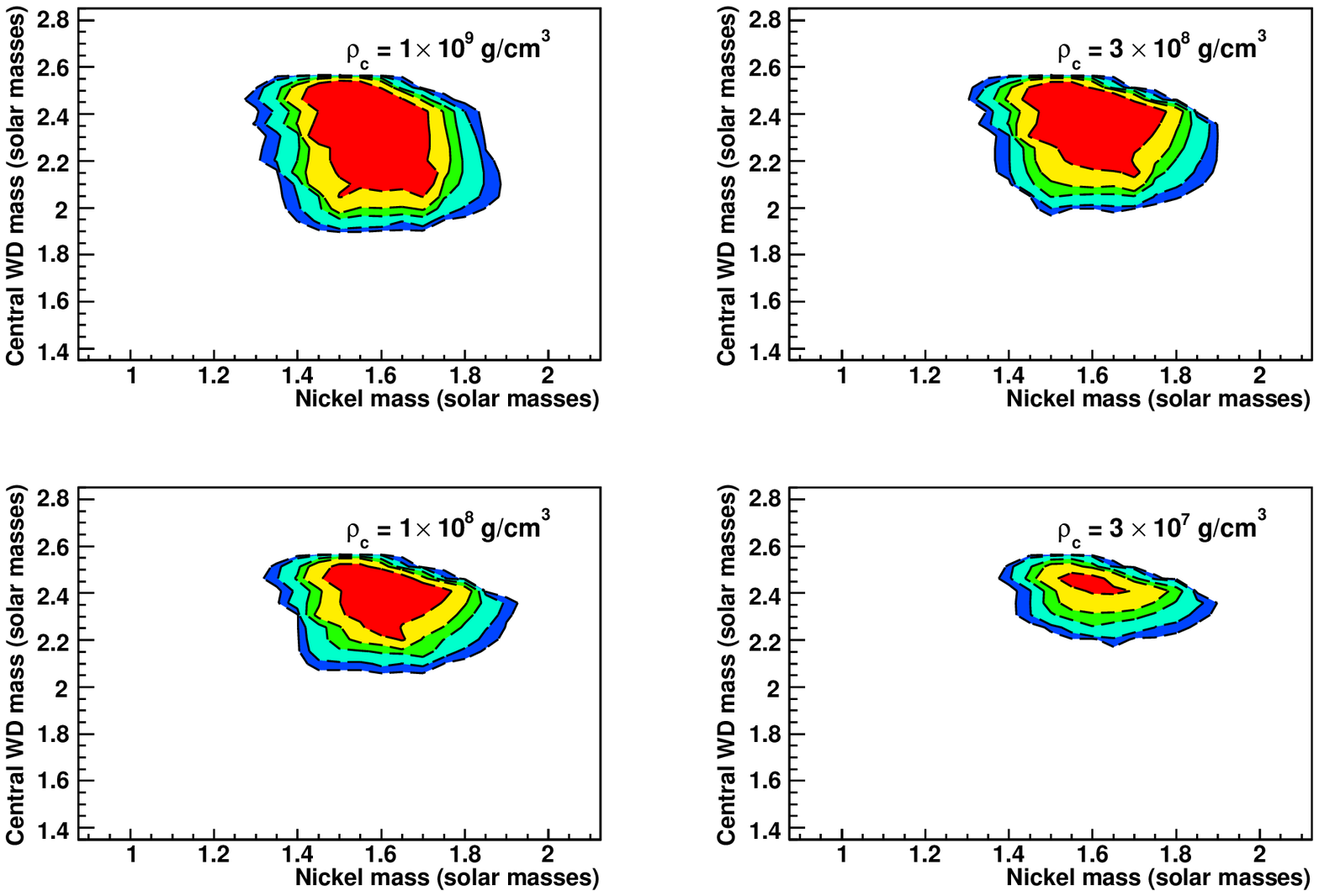}}
\caption{\small Constraints on progenitor model masses for SN~2007if,
assuming $\alpha = 1.3$
(see e.g. upper left panel of Figure~\ref{fig:cont-env})
for different possible central densities.
Probability density is normalized to that for the fiducial central density of
$\rho_c = 3 \times 10^9~\mathrm{g~cm^{-3}}$.
Colored regions are 68\% (red), 90\%, 95\%, 99\%, and 99.7\% CL (blue),
Top left:  $\rho_c = 1 \times 10^9~\mathrm{g~cm^{-3}}$.
Top right:  $\rho_c = 3 \times 10^8~\mathrm{g~cm^{-3}}$.
Bottom left:  $\rho_c = 1 \times 10^8~\mathrm{g~cm^{-3}}$.
Bottom right:  $\rho_c = 3 \times 10^7~\mathrm{g~cm^{-3}}$.}
\label{fig:cont-env-rcvar}
\end{figure*}

To this end, we repeat our analysis for $E(B-V) = 0$, $\alpha = 1.3$,
sampling a range of central densities from $3 \times 10^7~\mathrm{g~cm^{-3}}$
(close to that assumed for DET2ENVN) to our fiducial density.
We normalize the probability densities to our fiducial model as in
Section \ref{sec:reddening} above.
The results are shown in Figure \ref{fig:cont-env-rcvar}.
We find that although very low values of the
central density typical of DET2ENVN are disfavored, a wide range of possible
values are compatible with the data.  While the \nickel\ mass remains
unaffected, the total mass increases, and the envelope mass fraction needed
to decelerate the explosion to the observed velocities increases dramatically.
At central densities typical of DET2ENVN, an envelope of nearly 1.0 \Msol\
is needed to match the observations.

It is possible that other observables associated with our spectral time series
(for example, the detailed chemical composition of the ejecta) could provide
additional constraints on the central density.  Such detailed modeling is
beyond the scope of this paper, and awaits a refined theoretical description
of super-Chandrasekhar-mass SN~Ia progenitors, including hydrodynamic and
radiative transfer simulations of the resulting explosions.

% ############################################################################

\section{Are shells common in super-Chandrasekhar-mass explosions?}

\begin{deluxetable*}{lcccccccccc}
\label{tbl:superC-review}
\tabletypesize{\footnotesize}
\tablecaption{Comparison of super-Chandrasekhar-mass SN~Ia candidates}
\tablehead{
   \colhead{Attribute} &
   \multicolumn{2}{c}{SN~2003fg} &
   \multicolumn{2}{c}{SN~2006gz} &
   \multicolumn{2}{c}{SN~2007if} &
   \multicolumn{2}{c}{SN~2009dc} &
   \multicolumn{2}{c}{Population}  \\
   \colhead{} &
   \colhead{} &
   \colhead{} &
   \colhead{} &
   \colhead{} &
   \colhead{} &
   \colhead{} &
   \colhead{} &
   \colhead{} &
   \multicolumn{2}{c}{Statistics}
}
\startdata
m$_V$                              & \multispan{2}\hfil $20.43\pm0.05$ \hfil
                                   & \multispan{2}\hfil $15.99\pm0.01$\tablenotemark{b} \hfil
                                   & \multispan{2}\hfil $17.26\pm0.03$              \hfil
                                   & \multispan{2}\hfil $15.20\pm0.16$\tablenotemark{c}  \hfil &&\\[1mm]
%                                  & 15.06 is V originally measured from figure 1
%                                  & 14.98 is V remeasured from figure 1; then add 0.22 for MW
%                                  & 15.19 is quoted B & agrees with figure 1 minus MW
%                                  & then add 0.29 for MW to give Bmax = 15.48
Galactic E(B-V)                    & \multispan{2} 0.013 
                                   & \multispan{2} 0.063
                                   & \multispan{2} 0.079
                                   & \multispan{2} 0.071   &&\\[1mm]
% Distance modulii use H=71 and z_cmb for each galaxy
Distance Modulus                   & \multispan{2}\hfil $40.430\pm0.005$\tablenotemark{a}  \hfil
                                   & \multispan{2}\hfil $35.025\pm0.060$\tablenotemark{b} \hfil
                                   & \multispan{2}\hfil $37.569\pm0.028$              \hfil
                                   & \multispan{2}\hfil $34.859\pm0.051$\tablenotemark{c}  \hfil &&\\[1mm]
% M_V values correct for MW extinction using R_V = 3.1 and use the above DM values
M$_V$                              & \multispan{2}\hfil $-20.04\pm0.05$\tablenotemark{a}  \hfil
                                   & \multispan{2}\hfil $-19.23\pm0.06$\tablenotemark{b} \hfil
                                   & \multispan{2}\hfil $-20.39\pm0.04$              \hfil
                                   & \multispan{2}\hfil $-19.88\pm0.17$\tablenotemark{c}  \hfil &&\\[1mm]
% These are old M_V values as reported in the respective papers
%M$_V$                              & \multispan{2}\hfil $-19.93\pm0.06$\tablenotemark{a}  \hfil
%                                   & \multispan{2}\hfil $-19.19\pm0.09$\tablenotemark{b} \hfil
%                                   & \multispan{2}\hfil $-20.34\pm0.03$              \hfil
%                                   & \multispan{2}\hfil $-19.90\pm0.15$\tablenotemark{c}  \hfil &&\\[1mm]
% These Bmax-Vmax values are corrected for MW extinction; 03fg is based on SALTII 2
$B_{max}-V_{max}$                  & \multispan{2}\hfil $+0.14$ \hfil
                                   & \multispan{2}\hfil $+0.03$\tablenotemark{b} \hfil
                                   & \multispan{2}\hfil $+0.15$ \hfil
                                   & \multispan{2}\hfil $+0.21$ \hfil                  &&\\[1mm]
{$v_\mathrm{Si~II}$~(km~s$^{-1}$)} & \multispan{2}\hfil $\sim 8000$    \hfil            
                                   & \multispan{2}\hfil $\sim 13000$   \hfil                           
                                   & \multispan{2}\hfil $\sim 9000$    \hfil      
                                   & \multispan{2}\hfil $\sim 8000$    \hfil
                                   & \multispan{2}\hfil 9500~(2100) \hskip 0.45truein   \hfil     \\[1mm]
%B Stretch                          & \multispan{2}\hfil 1.13\tablenotemark{a} \hfil   
%                                   & \multispan{2}\hfil                       \hfil             
%                                   & \multispan{2}\hfil 1.15                  \hfil
%                                   & \multispan{2}\hfil                       \hfil  &&\\[1mm]
$\Delta$m$_{15}$(B)                & \multispan{2}\hfil 0.94\tablenotemark{c} \hfil   
                                   & \multispan{2}\hfil $0.69\pm0.04$\tablenotemark{b}   \hfil          
                                   & \multispan{2}\hfil $0.71\pm0.06$         \hfil
                                   & \multispan{2}\hfil $0.65\pm0.03$\tablenotemark{c}   \hfil    
                                   &\multispan{2}\hfil 0.75 (0.11) \hskip 0.45truein \hfil \\[1mm]
\ion{C}{2} 6580 persistance        & \multispan{2}\hfil not detected   \hfil            
                                   & \multispan{2}\hfil $+10$ ($< +11$) days \hfil                    
                                   & \multispan{2}\hfil $+7$ ($< +10$) days  \hfil
                                   & \multispan{2}\hfil $+6$ ($<+18$) days   \hfil
                                   & \multispan{2}\hfil  $+$8~(2)~days \hskip 0.45truein \hfil \\[1mm]
Host features                      & \multispan{2}\hfil emission lines;   \hfil         
                                   & \multispan{2}\hfil Scd spiral; no local      \hfil                       
                                   & \multispan{2}\hfil emission lines;   \hfil   
                                   & \multispan{2}\hfil S0; no star       \hfil       &&\\
                                   & \multispan{2}\hfil in tidal feature  \hfil      
                                   & \multispan{2}\hfil emission excess\tablenotemark{d}  \hfil
                                   & \multispan{2}\hfil dwarf                 \hfil
                                   & \multispan{2}\hfil formation?\tablenotemark{e} \hfil &&\\[1mm]
Host EW(Na~I~D)                    & \multispan{2}\hfil $<1.7$~\AA\ 95\% CL   \hfil
                                   & \multispan{2}\hfil $0.33\pm0.03$~\AA     \hfil           
                                   & \multispan{2}\hfil $<0.12$~\AA\ 95\% CL  \hfil
                                   & \multispan{2}\hfil 1.0~\AA\tablenotemark{c}   \hfil &&\\[1mm]
%$E(B-V)_\mathrm{Na~I~D}$           & \multispan{2}                          
%                                   & \multispan{2}                                          
%                                   & \multispan{2}                     
%                                   & \multispan{2}                           &&\\[1mm]
$E(B-V)_\mathrm{Lira}$             & \multispan{2}\hfil $0.21\pm0.04$   \hfil           
                                   & \multispan{2}\hfil $0.18\pm0.05$   \hfil                           
                                   & \multispan{2}\hfil $0.18\pm0.04$   \hfil      
                                   & \multispan{2}\hfil $0.37\pm0.08$   \hfil           &&\\[1mm]
\cutinhead{Quantities Corrected For Host Extinction Using \ion{Na}{1}~D\tablenotemark{f}}
                                   &  Shallow     &    Steep     
                                   &  Shallow     &    Steep     
                                   &  Shallow     &    Steep     
                                   &  Shallow     &    Steep
                                   &  Shallow     &    Steep    \\[1mm]
Corrected M$_V$                    &  $-20.04$    & $-20.04$  
                                   &  $-19.38$    & $-19.74$                   
                                   &  $-20.39$    & $-20.39$
                                   &  $-20.35$    & $-21.43$    
                                   & $-20.04$  ($0.40$)
                                   & $-20.40$  ($0.64$) \\[1mm]
Corrected $B_{max}-V_{max}$        &  $+0.14$     & $+0.14$    
                                   &  $-0.02$     & $-0.14$                    
                                   &  $+0.15$     & $+0.15$
                                   &  $+0.06$     & $-0.29$     
                                   & $+0.08$  ($0.07$)
                                   & $-0.03$  ($0.19$)  \\[1mm]
Corrected $E(B-V)_\mathrm{Lira}$   &  $+0.20$     & $+0.20$   
                                   &  $+0.13$     & $+0.02$                    
                                   &  $+0.18$     & $+0.18$
                                   &  $+0.22$     & $-0.13$     
                                   &  $+0.18$ ($0.03$)
                                   &  $+0.07$ ($0.13$)     \\

% Population stats calculated using NaD_corr.pro
%
% 03fg Lira colors:
% SN2003fg K-correction and Lira excess
% SN2003fg         @ 0.2440 : MegaCam g-r = 23.54  - 22.09  = 1.45  +/- 0.03
% SN2003fg MWcor   @ 0.2440 : MegaCam g-r = 23.54  - 22.09  = 1.44  +/- 0.03
% SN2007if 09_293  @ 0.2440 : MegaCam g-r = 20.704 - 19.534 = 1.170 +/- 0.024
% SN2007if 09_293  @ 0.0000 : Bessell B-V = 19.901 - 18.914 = 0.997 +/- 0.016
% SN2007if 09_293  @ 0.0000 : Andicam B-V = 19.871 - 18.934 = 0.937 +/- 0.020
% SN2003fg: K-cor  g-r --> B-V = 0.997 - 1.170 = -0.173
%           B-V = 1.44 - 0.173 = 1.17 +/- 0.04
%
% At day 40 Lira relation predicts B-V = 0.9610
%
% So, for SN2003fg day 40 Lira E(B-V) = 1.17 - 0.96 = 0.21 +/- 0.04 (stat)
% E(B-V) = 0.21 - 0.013 = 0.20 after correcting for MW

\enddata
\label{tbl:superC-review}
%\tablecomments{Comparison of observational properties among the various
%SNe~Ia claimed to have masses in excess of the Chandrasekhar limit.
%$v_\mathrm{Si~II}$, $B-V$ and $M_V$ are all measured or estimated at
%maximum light.  Reddening estimates from the Lira relation and from
%\ion{Na}{1}~D absorption are both given when applicable.}
\tablenotetext{a}{\cite{howell06}}
\tablenotetext{b}{\cite{hicken07}, with corrected distance modulus uncertainty.}
\tablenotetext{c}{\cite{yamanaka09}}
\tablenotetext{d}{There is no H$\alpha$ emission remaining in spectra of SN~2006gz after subtraction of an interpolated background.}
\tablenotetext{e}{Inspection of SDSS images shows uniformly red color.}
\tablenotetext{f}{For SN~2003fg and SN~2007if the most probable value, EW(\ion{Na}{1} D) = 0, was used}
\end{deluxetable*}

Our detailed observations of SN~2007if have produced a relatively specific
physical picture of the explosion, providing a framework within which
to interpret SN~2003fg-like SNe as a subclass.
A summary of the observational properties of the various
super-Chandrasekhar-mass SN~Ia candidates found to date is given in 
Table~\ref{tbl:superC-review}; we discuss what is known about these SNe below,
in the context of our findings. The data have been put on a uniform distance
scale, and measurements and uncertainties for some quantities have been
improved or corrected relative to the original sources employing the
techniques we have applied to SN~2007if.

\cite{maeda09b} argue that an aspherical explosion is needed to explain the
high luminosity of SN~2003fg, based on the relatively fast decay time of its
bolometric light curve ($t_{+1/2} = 13$~days) and low peak velocity,
and raise doubts about the high mass of the progenitor as interpreted
in the binding-energy framework of \cite{howell06}.  Conversely, they argue
that the blue observed color, high velocities, and long bolometric decay time
of SN~2006gz are consistent with a roughly spherical super-Chandrasekhar-mass
progenitor.

Our above analysis of SN~2007if allows us to address the criticism of
\cite{maeda09b} regarding the failures of a spherical geometry to describe
the properties of SN~2003fg-like SNe.  Their work assumed an essentially
undisturbed ejecta structure close to an exponential in velocity, in which
the peak \ion{Si}{2} velocity was a suitable proxy for the kinetic energy
of the explosion.  We have seen that, for density structures with shells,
this assumption is no longer true; powerful explosions with high kinetic
energy can have low-velocity, slowly-evolving photospheres.

A similar scenario may hold for SN~2003fg.  Although it is difficult to test
because of the limitations of the data, the existing observations of
SN~2003fg are broadly consistent with a SN~2007if-like scenario.
The single published spectrum shows a low velocity which may or may not
have been part of a plateau.  SALT2 fits to various
near-maximum subranges of the SNLS photometry result in values of $B-V$
at $B$-band maximum ranging between 0.03 and 0.14,
% $c = 0.22 \pm 0.03$, corresponding to $B-V = +0.15$ at $B$-band maximum
at odds with the published estimate $B-V = -0.15$ \citep{howell06,hicken07}.
However, the published (statistical only) error bars on $B-V$ from
\cite{howell06} probably underestimate the true error, given that there was
no rest-frame $B$-band data at maximum light for SN~2003fg
(Howell 2009, private communication).
A SALT2 fit to all available SNLS data gives $c = 0.22 \pm 0.03$,
or $B-V = +0.14$ at $B$-band maximum after correction for Galactic extinction;
this color is consistent with the color we obtain from synthetic photometry
of the published SN~2007if spectrum from SNLS \citep{howell06}.
While the evidence is far from conclusive, the spectral match to SN~2007if
near maximum light, and the arguments presented in \cite{maeda09b}, suggest
that SN~2003fg may be best interpreted as a super-Chandrasekhar-mass SN~Ia
with dense shell structure, like SN~2007if.

Likewise, SN~2009dc was also a very red SN with slow velocity
evolution.  \cite{yamanaka09} note that the very large value of the
reddening, $E(B-V)_\mathrm{host} = 0.37$, derived from the Lira relation
is at odds with the estimate $E(B-V)_\mathrm{host} = 0.15$ obtained from the
shallow TBC relation.  They adopt the latter value which, after correction,
produces $B-V \sim +0.2$ for SN~2009dc near peak.  This is broadly similar
to the intrinsic color we expect for SN~2007if.  We note that
\cite{yamanaka09} do not appear to correct for the efficiency factor $\alpha$
in their derivation of the \nickel\ mass of SN~2009dc; however, they also
assume a fairly typical 20-day rise time for the SN, when the true value
(not strongly constrained by pre-maximum observations) is probably higher,
given the SN's slow decline.  These two errors have opposite and comparable
effects on the derived \nickel\ mass, so it turns out that their estimate
should in fact be accurate, and hence comparable to SN~2003fg.
Taken together with the low, slowly-evolving
($\sim 50$~km~s$^{-1}$~d$^{-1}$ to day $+17$)
\ion{Si}{2} velocity, these observations are broadly compatible
with a shell-structure density profile for the ejecta of SN~2009dc.
The strong detection of \ion{C}{2} $\lambda 6580$ in the near-maximum-light
spectra of SN~2009dc, at comparable velocity to \ion{Si}{2}, suggests either
that a larger fraction of carbon remained unburned than in SN~2007if,
or that a pre-existing carbon/oxygen envelope may have been mixed into the
ejecta after burning was complete.

The fact that the Lira relation appears to overcorrect the reddening for
SN~2007if and SN~2009dc makes the interpretation of SN~2006gz problematic.
The supposed high luminosity and mass of SN~2006gz rests on the Lira relation,
corroborated by a low-significance detection of \ion{Na}{1} and the assumption
of a steep slope for the TBC relation, and the assumption $R_V = 3.1$.
The shallow TBC relation (adopted for SN~2009dc)
gives $E(B-V)_\mathrm{host} = 0.05$, which would make SN~2006gz somewhat
brighter and redder than average, but not unreasonably so for normal SNe~Ia.

To further examine this question we use the measurements compiled in
Table~\ref{tbl:superC-review} to more closely examine the properties of
this population under different extinction treatments. To aid in this
comparison we have remeasured \ewna\ from the host of SN~2006gz using the
same approach as for SN~2007if, resulting in a significant ($5\times$)
reduction in the uncertainty. By examining $M_V$,
$B_\mathrm{max}-V_\mathrm{max}$,
and the color excess, $E(B-V)_\mathrm{Lira}$, with respect to the Lira
relation we find that the properties of this population are most
uniform when the shallow TBC relation is used to correct for host-galaxy
extinction.  The resulting RMS in $M_V$ is only 0.40~mag, the RMS in
$B_{max}-V_{max}$ is only 0.07~mag, and the Lira color excess
has a remarkably small RMS of only 0.03~mag with this treatment.
Use of the steep TBC relation doubles or triples the RMS for each of these
quantities, suggesting that
it is disfavored by these data.  Using the shallow TBC \ewna\ extinction
correction places SN~2006gz at $M_V = -19.38$, well below the population
mean of $\langle M_V\rangle=-20.04$. (In fact, without SN~2006gz the mean
brightness of the population increases to $\langle M_V\rangle=-20.26$
and the RMS is only 0.16~mag.)

Other indirect evidence also points to a lower mass and intrinsic luminosity
for SN~2006gz.  Although the decay time of the SN~2006gz light curve is long,
the rise time is short, lowering the \nickel\ mass estimate relative to what
one might expect given the event's $\Delta m_{15}(B)$.  Late-time
spectroscopy and photometry of SN~2006gz \citep{maeda09a}
do not provide strong evidence for a large mass of \nickel\ in the explosion.
The \ion{Si}{2} velocity at maximum is typical of normal SNe~Ia, which
suggests neither a high binding energy nor a very massive shell.
While the plateau in the \ion{Si}{2} velocity for SN~2006gz supports the
existence of an envelope, as \cite{hicken07} suggest, the high velocity of
the supposed plateau (13000~km~s$^{-1}$) and its short duration, as well as the
blue color of the SN~relative to the SN~2003fg-like SNe, argue for an envelope
with mass only about 3\% of the white dwarf mass
(0.04 \Msol\ for $M_\mathrm{WD} = \mch$).
Thus, while SN~2006gz might have been a
DD merger and does show some evidence for a low-mass shell, it seems to be
different from the red, unambiguously overluminous SN~2003fg-like SNe.
If the short rise times of the PDD models of \cite{hk96} relative to other
explosion mechanisms are realized in nature, SN~2006gz may well have been
a Chandrasekhar-mass PDD; the models PDD9 and PDD1a compare favorably.

In short, comparison of SN~2007if with other events
suggests the following about SN~2003fg-like SNe (excluding SN~2006gz):
\begin{enumerate}
\item The tamped detonation scenario, in which the high-velocity ejecta
      are decelerated by an envelope, provides a cleaner explanation for
      the low photospheric velocities than a high binding energy alone,
      and it does so in the context of a nearly spherical explosion.
\item The presence of a shell also explains the long rise times and
      red colors of these events.
\item While the late-time color evolution of SN~2003fg-like SNe appears to
      have a slope similar to the Lira relation for normal SNe~Ia,
      the assumption that the intrinsic colors of these events at late
      times are similar to normal SNe~Ia probably leads to an overestimate
      of the reddening.
\end{enumerate}
More detailed analysis for these three SNe can be used to determine likely
values for the progenitor masses, allowing for the possibility of shells,
and (for tamped detonations) to constrain the mass of the envelope.
Such an exercise on a representative sample of SNe may be helpful in
refining theoretical models of the double-degenerate merger process.

% SAVE FOR DISCUSSION???
Finally, we note that as an extremely faint system, the host of SN~2007if is most likely of
low-mass and low-metallicity. The tidal tail in which SN~2003fg occurred
may have similar properties. In contrast, SN~2006gz occurred in a luminous
spiral and SN~2009dc occurred in an apparently quiesent elliptical galaxy.
Thus, it remains unclear what role the progenitor metallicity may play in
producing such events.

% ############################################################################

\section{Conclusions}

Our observations of SN~2007if provide detailed observations of the
evolution of a candidate super-Chandrasekhar-mass SN~Ia event,
allowing new constraints on their progenitors.  SN~2007if is the brightest
SN~Ia yet discovered, with an inferred \nickel\ mass of $1.6 \pm 0.1$~\Msol;
the near-IR light curves demonstrate evidence
for a large fraction of iron-peak elements in the ejecta, some of which may
have been mixed into the shell.  The slow \ion{Si}{2} velocity evolution near
maximum is inconsistent with ``normal'' SN~Ia photospheric evolution of
expanding ejecta with a nearly exponential density structure, and is more
readily interpreted as evidence for an overdense shell in the ejecta.
Further evidence for an unusual density structure in the ejecta comes from
the late-time bolometric light curve, which implies a higher gamma-ray escape
fraction than one might expect for a progenitor with normal photospheric
evolution.  Interpreting the observations in the context of a
tamped detonation model representative of a double-degenerate merger
\citep[as in the DET2ENVN models of][]{kmh93},
we use the SN~2007if bolometric light curve to establish the first
constraint on the total mass for a super-Chandrasekhar-mass SN~Ia progenitor.
Our mass estimate should strictly be construed as a lower bound,
since reddening by dust and a low central density will both result in a
higher mass; however, extreme compositions and unrealistically
high masses, at the edge of the allowed parameter space,
result if either the reddening or the central density are
very different from those used in our fiducial analysis in
Section \ref{sec:confint-fiducial}.

Better models of the progenitors and explosions of SN~2003fg-like SNe~Ia
are urgently needed,
mainly because of the theoretical limits on the existence of supermassive
white dwarfs \citep{piro08,chenli09} which provide possible initial
conditions for a SN~2003fg-like explosion.  In our work, we nevertheless assume
the existence of a central supermassive white dwarf and use the binding energy
formula of \cite{yl05}.  This formula has been validated only for
$M < 2.0$ \Msol, but other works have performed similar extrapolations
\citep[e.g.,][]{jbb06} simply because no appropriate alternative exists
for modeling hypothetical explosions of supermassive white dwarfs.

Further uncertainty relates to the properties of the carbon/oxygen envelope.
The envelope considered in DET2ENVN and in this work is relatively small in
extent ($R \sim 10^{10}$ cm), so the shock interaction which produces the
shell can be approximated as occurring instantaneously, with the ejecta in
homologous expansion thereafter.  Our simple modeling suggests that a large
fraction of the kinetic energy
(up to 20\% of the initial value, or a few $10^{50}$ erg)
must be dissipated in the transition from explosion to the final state with
fully developed shell.  If the envelope is more diffuse, the shock
interaction and conversion of kinetic energy would proceed over a longer
period of time, with some of the energy released at optical wavelengths.
The interaction could produce a ``pseudo-continuum'' similar to those seen
in SNe IIa \citep{hamuy03,mwv04,snf2005gj,sdss2005gj}, without the telltale
H and He emission, but possibly including C and O emission, toplighting
absorption from those elements in the SN atmosphere \citep{branch00}.
This could explain the blue, relatively featureless continuum seen
in SN~2007if at $-9$~days.  It would also contribute to the long rise time
and high luminosity of the event, thus inflating our derived \nickel\ mass.
Although we find this scenario unlikely, very early-phase observations of
future events similar to SN~2007if, along with theoretical calculations of
synthetic spectra of such an interaction, will be needed in order to
constrain it.  However, even if our \nickel\ mass is too high, the uniformity
of the optical spectra at late times provide some evidence that the
interaction produces negligible late-time luminosity.
The lower limit on the total ejected mass from the late-time bolometric
light curve should therefore remain secure, since the anticorrelation
shown in Table~\ref{tbl:mlim-alpha-vary} and Figure~\ref{fig:cont-env-run2}
requires higher total masses for models with lower \nickel\ mass.

More definitive statements about the progenitor mass and observational
properties await detailed radiative transfer calculations.  These will require
more accurate density structures of super-Chandrasekhar-mass progenitors,
starting from appropriate explosion models --- which may in turn require
initial conditions appropriately chosen, e.g., from simulations of the dynamics
of double-degenerate inspiral events.
The theoretical uncertainty on the parameter $\alpha$ relating the
radioactivity and bolometric luminosities at maximum light presented another
limiting factor, in light of the degeneracy between the \nickel\ mass and
the total mass for a particular bolometric
light curve.  Clearer predictions for the rise time and intrinsic color
of super-Chandrasekhar-mass SNe Ia, with and without envelopes or
shell structure, would also be helpful in understanding these events.

A larger sample of SN~2003fg-like SNe is also needed to assess whether velocity
plateaus truly are generic in these objects; this will require fairly
intensive spectroscopic follow-up as quickly after explosion as possible.
Our extensive spectral time series, including one of the earliest
(the earliest if SN~2006gz is not included in this subclass) known spectrum
of a SN~2003fg-like SN~Ia ($-9$~days), should aid in the prompt identification
of these events for future follow-up, especially in cases where redshifts of
faint host galaxies are difficult to obtain.
Optical/NIR light curves out to 100~days past maximum light would be useful
to confirm the high mass in these explosions.  NIR observations could be
particularly useful for reducing the uncertainty in the dust extinction
(and thereby testing the Lira relation) for SN~2003fg-like SNe, and for
determining the distribution of iron-peak elements for these objects.
Modeling of the photospheric \citep{mazzali07,mazzali08,agy09}
and nebular \citep[e.g.,][]{mazzali97,maeda09a} spectral time series
would be useful to provide independent constraints on the \nickel\ mass,
the total ejected mass, and the specific products of nuclear burning in
super-Chandrasekhar-mass SNe, allowing us to learn more about the details
of the explosion.

These detailed results will allow us to begin to connect SN~2003fg-like SNe
to potential counterparts among ``normal'' SNe Ia.
Construction of late-time bolometric light curves for a representative
sample of less luminous velocity-plateau and/or LVG \citep{benetti05}
SNe Ia could enable new measurements of the ejected masses and shell mass
fractions for these objects, presenting strong new constraints on their
progenitors.  Comparison of these SNe with SN~2003fg-like SNe could also
enable observational methods to isolate or ``tag'' a subclass demanding
different treatment when measuring luminosity distances.
\cite{qhw07} caution that shell-structure SNe should be treated differently
from normal SNe~Ia in cosmology fits, having intrinsically redder colors
and violating the typical width-luminosity relations.  \cite{howell06} give
a similar warning relating to a possible class of super-Chandrasekhar-mass
progenitors with smaller inferred masses than SN~2003fg.  Limiting the
impact of such scenarios is crucial for the success of future
(Stage III \& IV) dark energy experiments. Fortunately, with the addition
of SN~2007if, the properties of this sub-class are becoming more clear, and
their clear spectral peculiarity establishes a means to recognize such
events.

% ############################################################################

\acknowledgments

The authors are grateful to the technical and scientific staffs of the
University of Hawaii 2.2~m telescope, the W. M. Keck Observatory,
and Palomar Observatory, to the QUEST-II collaboration, and to HPWREN for
their assistance in obtaining these data.
Some of the data presented herein were obtained at the W. M. Keck
Observatory, which is operated as a scientific partnership among the
California Institute of Technology, the University of California,
and the National Aeronautics and Space Administration.  The Observatory
was made possible by the generous financial support of the W. M. Keck
Foundation.
The authors wish to recognize and acknowledge the very significant cultural
role and reverence that the summit of Mauna Kea has always had within the
indigenous Hawaiian community.  We are most fortunate to have the opportunity
to conduct observations from this mountain.
This work was supported by the Director, Office of Science, Office of
High Energy Physics, of the U.S. Department of Energy under Contract No.
DE-AC02-05CH11231; by a grant from the Gordon \& Betty Moore Foundation;
and in France by support from CNRS/IN2P3, CNRS/INSU, and PNC.
RS acknowledges support from National Science foundation grant 0407297.
YC acknowledges support from a Henri Chretien International Research 
Grant administrated by the American Astronomical Society, and from the 
France-Berkeley Fund.
A.G.-Y. is supported by the Israeli Science Foundation, an EU Seventh
Framework Programme Marie Curie IRG fellowship and the Benoziyo Center for
Astrophysics, Minerva program, a research grant from the Peter and Patricia
Gruber Awards, and the William Z. and Eda Bess Novick New Scientists Fund at
the Weizmann Institute.
This research used resources of the National Energy Research Scientific
Computing Center, which is supported by the Director, Office of Science,
Office of Advanced Scientific Computing Research, of the U.S. Department
of Energy under Contract No. DE-AC02-05CH11231.  We thank them for a generous
allocation of storage and computing time.
HPWREN is funded by National Science Foundation Grant Number ANI-0087344,
and the University of California, San Diego.
IRAF is distributed by the National Optical Astronomy Observatories,
which are operated by the Association of Universities for Research
in Astronomy, Inc., under cooperative agreement with the National
Science Foundation.
The spectra of SN~1999ee were obtained through the SUSPECT Supernova
Spectrum Archive, an online database maintained at the University of
Oklahoma, Norman.
We thank Andy Howell for providing photometry and the near-maximum spectrum
of SN~2003fg, and Masayuki Yamanaka for providing \ion{Si}{2} and \ion{C}{2}
velocity measurements derived from SN~2009dc spectra.
The SMARTS 1.3m observing queue receives support from NSF grant AST-0707627.
We thank John Holtzman and Jon Cough for their assistance in photometrically
monitoring our SN candidates with the NMSU 1~m telescope.
We thank Dan Birchall for his assistance in collecting data with SNIFS,
and for his helpful commentary on, and proofreading of, the manuscript.
We also thank Charles Bailyn, Richard Pogge, and Kevin Krisciunas for their
assistance in characterizing the ANDICAM system throughput, and Daniel Kasen
and Alan Calder for helpful discussions.

{\it Facilities:}
\facility{UH:2.2m ()},
\facility{Hale ()},
\facility{PO:1.2m ()},
\facility{CTIO:1.5m ()}

% ############################################################################

% \section*{References}
% 
% \begin{itemize}
% \item[{[1]}] Press, W. H. et al.  {\itshape Numerical Recipes.}
%    Cambridge University Press, 1992.
% \end{itemize}

% >>>>>>>>>>>>>>>>>>>>>>>>>>>>>>>>>>>>>>>>>>>>>>>>>>>>>>>>>>>>>>>>>>>>>>>>>>>>

% To process this document or documents that use this as a template:
%
% 1.  Run Latex on it.  Look at the page numbers.  Edit the page number
%     quantity at the top of the file ``{\tt of 1}''.
% 2.  Re-run Latex on it.
% 3.  Run dvips with the -p 1 option to skip the title page.
%
% Note: if you have postscript files to include in the document, they
% must be encapsulated postscript (.eps).  Furthermore, to use psfig,
% you need to keep a copy of psfig.tex in the directory where you are
% \LaTeX-ing your document.

\end{document}